\documentclass[12pt]{article}
\usepackage[utf8]{inputenc}
\usepackage{natbib}

\usepackage{etoolbox}

\patchcmd{\thebibliography}{\bibsection}{\section*{References}}{}{}
\patchcmd{\thebibliography}{\section*{\refname}}{\section*{References}}{}{}
\patchcmd{\thebibliography}{\chapter*{\bibname}}{\section*{References}}{}{}

\usepackage{setspace}
\usepackage[english]{babel}
\usepackage{graphicx}
\usepackage{indentfirst}
\usepackage{framed}
\usepackage[normalem]{ulem}
\usepackage{amsmath}
\usepackage{amsthm}
\usepackage{amssymb}
\usepackage{amsfonts}
\usepackage{dcolumn}
\usepackage{enumerate}
\usepackage{float}
\usepackage{tabularx}
\usepackage{caption}
\usepackage{subcaption}
\usepackage{multirow}
\usepackage[utf8]{inputenc}
\usepackage{tabularx}
\usepackage{array}
\usepackage{float}
\usepackage{hyperref}
\usepackage{tikz}
\usepackage{pgf}
\usepackage[nottoc,numbib]{tocbibind}
\usepackage{booktabs}

\usepackage[top=1 in,bottom=1in, left=1 in, right=1 in]{geometry}

\newcolumntype{L}{>{\centering\arraybackslash}m{3cm}}
\newcommand{\X}{\boldsymbol{X}}

\theoremstyle{definition}

\newtheorem{proposition}{Proposition}

\newcolumntype{Y}{>{\centering\arraybackslash}X}

\begin{document}

\begin{center}
\Large{Adaptive CUSUM Chart for Real-Time Monitoring of Bivariate Event Data}
\end{center}

\begin{center}
{\large Gokul Parakulum and Jun Li$^*$}\\
     Department of Statistics, University of California - Riverside\\
     $^*$Email: jun.li@ucr.edu.
\end{center}

\begin{abstract}
Monitoring time-between-events (TBE) data---where the goal is to track the time between consecutive events---has important applications across various fields. Many existing schemes for monitoring multivariate TBE data suffer from inherent delays, as they require waiting until all components of the observation vector are available before making inferences about the process state. In practice, however, these components are rarely recorded simultaneously. To address this issue, \cite{Zwetsloot_MainPaper} proposed a Shewhart  chart for bivariate TBE data that updates the process status as individual observations arrive. However, like most Shewhart-type charts, their method evaluates the process based solely on the most recent observation and does not incorporate historical information. As a result, it is ineffective in detecting small to moderate changes. To overcome this limitation, we develop an adaptive CUSUM chart that updates with each incoming observation while also accumulating information over time. 
Simulation studies and real data applications demonstrate that our method substantially outperforms the Shewhart chart proposed by \cite{Zwetsloot_MainPaper}, offering a robust and effective tool for real-time monitoring of bivariate TBE data.

\end{abstract}

{\bf Key words:} Adaptive CUSUM chart; Real-time monitoring;
Statistical process monitoring; Time-between-events

\section{Introduction}

Statistical Process Control (SPC) has traditionally focused on monitoring quality characteristics that follow Gaussian distributions. However, the shift in modern industries toward high-yield, high-reliability systems---characterized by extremely low defect rates---has necessitated the development of alternative monitoring strategies. In such contexts, conventional Shewhart, Cumulative Sum (CUSUM), and Exponentially Weighted Moving Average (EWMA) charts may be inadequate for detecting rare but critical events. This limitation has led to the adoption of Time-Between-Events (TBE) charts, which monitor the intervals between specified events. TBE charts are particularly well suited for processes where nonconformities are infrequent but consequential  (\cite{Ozsan2009, Xie.Textbook.2002, Xie_Goh_2011_TwoMEWMA_Gumbel}).

While univariate TBE charts have been extensively studied and applied---for example, \cite{Sparks2019, Sparks2020}, \cite{Ozsan2009}---many real-world systems involve multiple interrelated components whose events or failures are not independent. Monitoring each component separately using univariate charts can overlook essential dependence structures. For instance, in circuit board manufacturing, multiple components are often soldered in close physical proximity. Poor soldering conditions can introduce shared vulnerabilities that cause correlated failures across components. Monitoring each component separately would fail to capture such dependencies (\cite{Costa_2008}).

To address this limitation, multivariate TBE charts have been proposed. In the context of bivariate monitoring, \cite{Costa_2008} evaluated three practical approaches for extending univariate TBE charts to the bivariate setting: monitoring the sum, difference, or maximum of the two component event times. Their empirical analysis suggested that monitoring the difference between event times was the most effective among these three strategies.

A more comprehensive class of techniques, known as vector-based schemes, has also been developed. One notable example is the Multivariate Exponentially Weighted Moving Average (MEWMA) chart proposed by \cite{Xie_Goh_2011_TwoMEWMA_Gumbel}, which employs a multivariate monitoring statistic that accounts for the covariance structure among component event times. However, a key limitation of these methods is that they require all components of the multivariate vector to be observed before updating the chart. In systems where component event times are recorded asynchronously, this requirement can lead to detection delays that reduce
responsiveness and undermine the goal of early intervention. Below is a useful illustration from \cite{Zwetsloot_MainPaper} that helps visualize this issue:

\begin{figure}[H]
\centering
\includegraphics[width=0.75\textwidth]{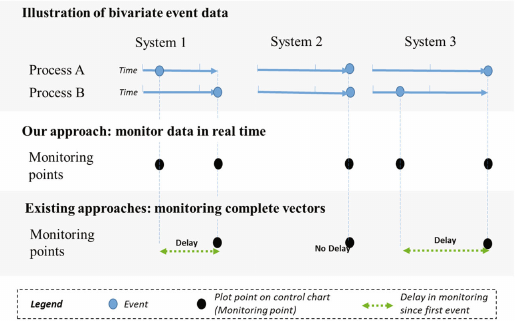}
\caption{Diagram demonstrating the aforementioned delay when waiting to observe both components shown in \cite{Zwetsloot_MainPaper}}
\end{figure}


Recognizing the challenge of asynchronous observations in bivariate settings, \cite{Zwetsloot_MainPaper} proposed a real-time Shewhart chart that updates the monitoring statistic immediately upon observation of either component, without waiting for the full observation vector. This framework represents a significant advancement for applications where time lags in data collection or failure reporting are unavoidable. However, as with most Shewhart-type charts, their method evaluates the process based solely on the most recent observation and does not incorporate past observations. Consequently, it lacks sensitivity to small or moderate changes. 

To overcome this limitation, we propose a real-time adaptive CUSUM chart specifically designed for bivariate asynchronous monitoring. Like the method of \cite{Zwetsloot_MainPaper}, our approach updates as soon as individual component observations arrive, thereby avoiding the delays inherent in vector-based methods. However, by leveraging the adaptive CUSUM framework---well known for its effectiveness in detecting small to moderate shifts (\cite{Page1954})---our method offers a more sensitive alternative. Across nearly all evaluated scenarios, our proposed adaptive CUSUM chart significantly outperforms the Shewhart chart introduced by \cite{Zwetsloot_MainPaper}. Thus, it provides a reliable and efficient solution for real-time monitoring of bivariate TBE data.

The remainder of this paper is organized as follows.  
Section~2 introduces the problem formulation and describes the transformations used to convert the real-time bivariate TBE monitoring problem into a univariate monitoring problem. Section~3 presents the proposed adaptive CUSUM chart for the transformed data.  
Section~4 reports simulation studies that evaluate the performance of the proposed procedure.  
Section~5 demonstrates the application of our method to a real dataset. Finally, Section~6 concludes with some closing remarks. All proofs are deferred to the Appendix.

\section{Problem Formulation and Data Transformation}\label{sec:preliminary}
\subsection{Real-Time Monitoring Framework}\label{sec:formulation}

Following \cite{Zwetsloot_MainPaper},  we consider a sequence of bivariate random variables \(\X_i = (X_{i,1}, X_{i,2})'\), $i=1,2,...$, where $X_{i,1}$ denotes the time to an event in the first subprocess and $X_{i,2}$ denotes 
the time to an event in the second subprocess. For example, \(X_{i,1}\) and \(X_{i,2}\) may represent time to blindness in the left and right eyes, respectively. In the context of circuit board manufacturing, they may denote the failure times of two components on the same board. Similar to \cite{Zwetsloot_MainPaper}, we assume that \(\X_{i}\) and \(\X_{j}\) are independent for all \(i \neq j\). 

Initially, the process is in control, meaning that $\X_i \sim F_0(x_1, x_2)$, where \(F_0\) is the in-control (IC) cumulative distribution function (CDF). After some unknown change-point time $\tau$,  the process shifts out of control, and $\X_i \sim F_1(x_1, x_2)$, where \(F_1\) is the out-of-control (OC) CDF. Thus, $\X_i$ follows the change-point model:
\begin{equation}
\label{eqn:CPM}
\X_i \sim
\begin{cases}
   F_0(x_1, x_2), & \text{if } i \leq \tau \\    
   F_1(x_1, x_2), & \text{if } i > \tau
\end{cases}
\end{equation}
As is standard in Phase II SPC, we assume that $F_0$ is either known or can be accurately estimated from a Phase I training dataset. 

Since \(X_{i,1}\) and \(X_{i,2}\) both represent event times with no inherent order---either one may occur first---we define $X_{i,(1)} \leq X_{i,(2)}$ as the order statistics of \(X_{i,1}\) and \(X_{i,2}\). That is,  $X_{i,(1)}$ denotes the earlier of the two event times,  and $X_{i,(2)}$ the later. In the applications considered in this paper, we assume that  observations arrive sequentially in the order $X_{1,(1)}$, $X_{1,(2)}$, $X_{2,(1)}$, $X_{2,(2)}$,...,$X_{i,(1)}$, $X_{i,(2)}$,..., over time. Our goal is to develop a real-time monitoring procedure capable of detecting distributional changes as quickly as possible based on these sequentially observed event times. In particular, we require the monitoring  statistic to update immediately upon the arrival of either $X_{i,(1)}$ or $X_{i,(2)}$ without waiting for the full bivariate vector $\X_i$. This can be achieved by applying a suitable univariate control chart to track these observations in real time. 
 
Among the various control charts available, the CUSUM chart is widely regarded for its effectiveness in detecting small to moderate shifts, due to its optimality properties (\cite{Moustakides1986}). However, traditional CUSUM charts assume independence among observations. In our setting, while the observations from different vectors---$(X_{i_1,(1)},X_{i_1,(2)})$ and $(X_{i_2,(1)},X_{i_2,(2)})$ for $i_1 \neq i_2$---are independent by assumption, the observations $X_{i,(1)}$ and $X_{i,(2)}$ within the same vector are not. This dependence arises both from the potential dependence between \(X_{i,1}\) and \(X_{i,2}\), and from the ordering operation itself---even if \(X_{i,1}\) and \(X_{i,2}\) are independent, their order statistics $X_{i,(1)}$ and $X_{i,(2)}$ are dependent. 

To address the challenge posed by the dependence between $X_{i,(1)}$ and $X_{i,(2)}$, we first propose a transformation to decorrelate those observations. 

\subsection{Transformation for Decorrelation} \label{sec:decorrelation}
We first define the variable \(V_i\) as:
\begin{equation}
\label{eqn:Vi}
V_i = \begin{cases}
   0, & \text{if } X_{i,1} < X_{i,2} \\    
   1, & \text{if } X_{i,1} > X_{i,2} \\
   2, & \text{if } X_{i,1} = X_{i,2}
\end{cases}
\end{equation}
In two of the three distributions considered later in the paper, \(\mathbb{P}(X_{i,1} = X_{i,2}) > 0\), so the case \(V_i = 2\) is non-trivial.  Based on this definition, the value of \(V_i\) indicates which component of $\X_i$ arrives first. Thus, whenever the first observation of $\X_i$ arrives, the value of  \(V_i\) is immediately known. For example, suppose $(X_{i,1}, X_{i,2})'=(3,2)'$. In this case, we first observe \(X_{i,(1)} = 2\), so  \(V_i = 1\), indicating that the first observation comes from the second component, and the second observed value, \(X_{i,(2)} = 3\), comes from the first component. In the special case where \(X_{i,1} = X_{i,2}\) (i.e., \(V_i = 2\)), we treat the entire vector as a single univariate observation.

Since $X_{i,(1)}$ and $X_{i,(2)}$ are correlated, we aim to decorrelate them. Assuming that $\X_i$ is in control, i.e., $\X_i \sim F_0$, we define $F^{(0)}_{X_{(1)}|V}$ as the conditional CDF of $X_{i,(1)}$ given $V_i$, and $F^{(0)}_{X_{(2)}|X_{(1)},V}$ as the conditional CDF of $X_{i,(2)}$ given $X_{i,(1)}$ and $V_i$. Both conditional CDFs can be derived from the IC distribution $F_0$, which is independent of the index $i$, so we omit the subscript \(i\) in their notation. Three widely used examples of $F_0$ will be provided later, along with their corresponding conditional CDFs $F^{(0)}_{X_{(1)}|V}$  and $F^{(0)}_{X_{(2)}|X_{(1)},V}$. 

We now define the following transformed variables:
\begin{equation}
\label{eqn:transform}
U_{i,(1)} = F^{(0)}_{X_{(1)}|V}(X_{i,(1)} \mid V_i), \qquad U_{i,(2)} = F^{(0)}_{X_{(2)}|X_{(1)},V}(X_{i,(2)} \mid X_{i,(1)}, V_i).
\end{equation}
Based on this transformation, we have the following result:

\begin{proposition}
\label{prop:IC}
When the original bivariate vector $\X_i$ is drawn from the IC distribution  $F_0$, the transformed variables \(U_{i,(1)}\) and \(U_{i,(2)}\) defined in (\ref{eqn:transform}) are independent and uniformly distributed on \([0,1]\) (denoted by $\text{Unif}[0,1]$). 
\end{proposition}

Based on Proposition~\ref{prop:IC}, applying the transformation in (\ref{eqn:transform}) to the original sequence $X_{1,(1)}$, $X_{1,(2)}$, $X_{2,(1)}$, $X_{2,(2)}$,...,$X_{i,(1)}$, $X_{i,(2)}$,... results in a new sequence $U_{1,(1)}$, $U_{1,(2)}$, $U_{2,(1)}$, $U_{2,(2)}$, ...,$U_{i,(1)}$, $U_{i,(2)}$, ..., all of which are i.i.d. $\text{Unif}[0,1]$ under the IC process. This transformation establishes $\text{Unif}[0,1]$ as the IC distribution of the transformed variables. To implement a classic CUSUM chart, both the IC and OC distributions must be specified. Therefore, in the following subsection, we study the OC distributions of the transformed variables $U_{i,(1)}$ and $U_{i,(2)}$. 

\subsection{OC Distribution Derivation} \label{sec:OCdist}

Since the transformation defined in (\ref{eqn:transform}) depends on the IC distribution \(F_0\), we follow \cite{Zwetsloot_MainPaper} and study OC distributions of the transformed variables \(U_{i,(1)}\) and \(U_{i,(2)}\) under three distribution families commonly used to model bivariate TBE data.

\subsubsection{Marshall–Olkin Bivariate Exponential (MOBE) Distribution}

The MOBE distribution, introduced by \cite{Marshall1967}, provides a tractable and interpretable bivariate extension of the exponential distribution. Specifically, if two random variables \(X_1\) and \(X_2\) jointly follow a MOBE distribution with parameters \(\lambda_1\), \(\lambda_2\), and \(\lambda_3\) (denoted by \(\text{MOBE}(\lambda_1, \lambda_2, \lambda_3)\)), its joint survival function is given by:
\[
S(x_1, x_2) = \mathbb{P}(X_1 > x_1, X_2 > x_2) = \exp\left(-\lambda_1 x_1 - \lambda_2 x_2 - \lambda_3 \cdot \max(x_1, x_2)\right).
\]

We assume that the IC distribution of \(\X_i\) is \(\text{MOBE}(\lambda_1,\lambda_2,\lambda_3)\), and define \(\Lambda := \lambda_1 + \lambda_2 + \lambda_3\). Based on results from \cite{Zwetsloot_MainPaper}, we obtain explicit expressions for the conditional distribution functions \(F^{(0)}_{X_{(1)}|V}\) and \(F^{(0)}_{X_{(2)}|X_{(1)},V}\) used in the transformation (\ref{eqn:transform}):
\begin{equation}
\label{eqn:IC1_MOBE}
F^{(0)}_{X_{(1)}|V}(x \mid v) = 1 - \exp(-\Lambda x), \qquad \text{for } v \in \{0,1,2\},
\end{equation}
\begin{equation}
\label{eqn:IC2_MOBE}
F^{(0)}_{X_{(2)}|X_{(1)},V}(y \mid x, v) = \begin{cases}
1 - \exp[(\lambda_2 + \lambda_3)(x - y)], & \text{if } v = 0, \\
1 - \exp[(\lambda_1 + \lambda_3)(x - y)], & \text{if } v = 1.
\end{cases}
\end{equation}

Applying the transformation in (\ref{eqn:transform}) using the conditional distributions in (\ref{eqn:IC1_MOBE}) and (\ref{eqn:IC2_MOBE}), the transformed variables \(U_{i,(1)}\) and \(U_{i,(2)}\) are i.i.d. \(\text{Unif}[0,1]\) under the IC process. The following proposition characterizes the OC distribution of the transformed variables when \(\X_i\) follows a shifted MOBE distribution.

\begin{proposition} \label{prop:OC_MOBE}
Suppose \(\X_i\) follows an OC distribution \(\text{MOBE}(\gamma_1, \gamma_2, \gamma_3)\), and define \(\Gamma := \gamma_1 + \gamma_2 + \gamma_3\). Then the transformed variables \(U_{i,(1)}\) and \(U_{i,(2)}\) are independent. Furthermore, their CDFs are given by:
\[
F_{U_{i,(1)}}(u) = 1 - (1 - u)^{\Gamma / \Lambda},
\]
and
\[
F_{U_{i,(2)}}(u) = \begin{cases}
1 - (1 - u)^{(\gamma_2 + \gamma_3) / (\lambda_2 + \lambda_3)}, & \text{if } V_i = 0, \\
1 - (1 - u)^{(\gamma_1 + \gamma_3) / (\lambda_1 + \lambda_3)}, & \text{if } V_i = 1.
\end{cases}
\]
\end{proposition}

\subsubsection{Marshall–Olkin Bivariate Weibull (MOBW) Distribution}

The MOBW distribution is a natural extension of the MOBE distribution, also briefly discussed in \cite{Marshall1967}. Specifically, if two random variables \(X_1\) and \(X_2\) jointly follow a MOBW distribution with parameters \(\lambda_1\), \(\lambda_2\), \(\lambda_3\), and \(\eta\) (denoted by \(\text{MOBW}(\lambda_1,\lambda_2,\lambda_3,\eta)\)), its joint survival function is given by:
\[
S(x_1, x_2) = \mathbb{P}(X_1 > x_1, X_2 > x_2) = \exp\left(-\lambda_1 x_1^\eta - \lambda_2 x_2^\eta - \lambda_3 \cdot \max(x_1, x_2)^\eta\right).
\]

We assume that the IC distribution of \(\X_i\) is \(\text{MOBW}(\lambda_1,\lambda_2,\lambda_3,\eta)\), and define \(\Lambda := \lambda_1 + \lambda_2 + \lambda_3\). Following \cite{Zwetsloot_MainPaper}, we obtain the conditional distributions \(F^{(0)}_{X_{(1)}|V}\) and \(F^{(0)}_{X_{(2)}|X_{(1)},V}\) used in the transformation (\ref{eqn:transform}):
\begin{equation}
 \label{eqn:IC1_MOBW}   
F^{(0)}_{X_{(1)} \mid V}(x \mid v) = 1 - \exp(-\Lambda x^\eta), \quad \text{for } v \in \{0, 1, 2\},
\end{equation}
\begin{equation}
 \label{eqn:IC2_MOBW} 
F^{(0)}_{X_{(2)} \mid X_{(1)}, V}(y \mid x, v) =
\begin{cases}
1 - \exp\left[(\lambda_2 + \lambda_3)(x^\eta - y^\eta)\right], & \text{if } v = 0, \\
1 - \exp\left[(\lambda_1 + \lambda_3)(x^\eta - y^\eta)\right], & \text{if } v = 1.
\end{cases}
\end{equation}

Applying the transformation in (\ref{eqn:transform}) using the conditional distributions in (\ref{eqn:IC1_MOBW}) and (\ref{eqn:IC2_MOBW}) yields transformed variables \(U_{i,(1)}\) and \(U_{i,(2)}\) that are i.i.d. \(\text{Unif}[0,1]\) under the IC process. Similar to \cite{Zwetsloot_MainPaper}, we assume that under the OC process only the scale parameters \(\lambda_1, \lambda_2, \lambda_3\) may change, while the shape parameter \(\eta\) remains constant. The OC distributions of the transformed variables are summarized in the following proposition.

\begin{proposition} \label{prop:OC_MOBW}
Suppose \(\X_i\) follows an OC distribution \(\text{MOBW}(\gamma_1, \gamma_2, \gamma_3,\eta)\), and define \(\Gamma := \gamma_1 + \gamma_2 + \gamma_3\). Then the transformed variables \(U_{i,(1)}\) and \(U_{i,(2)}\) are independent. Furthermore, their CDFs are given by:
\[
F_{U_{i,(1)}}(u) = 1 - (1 - u)^{\Gamma / \Lambda},
\]
and
\[
F_{U_{i,(2)}}(u) = \begin{cases}
1 - (1 - u)^{(\gamma_2 + \gamma_3)/(\lambda_2 + \lambda_3)}, & \text{if } V_i = 0, \\
1 - (1 - u)^{(\gamma_1 + \gamma_3)/(\lambda_1 + \lambda_3)}, & \text{if } V_i = 1.
\end{cases}
\]
\end{proposition}

\subsubsection{Gumbel Distribution}

The Gumbel distribution is a well-known bivariate distribution that also features univariate exponential marginals. Using the same notation as presented in \cite{Zwetsloot_MainPaper}, the survival function of a Gumbel distribution (denoted by \(\text{Gumbel}(\theta_1, \theta_2, \delta)\)) is given by:
\[
S(x_1, x_2) = \exp\left(-C(x_1, x_2)^\delta\right),
\]
where \( C(x_1, x_2) = \left(x_1/\theta_1\right)^{1/\delta} + \left(x_2/\theta_2\right)^{1/\delta} \), and \(\theta_1, \theta_2, \delta > 0\).

We assume that the IC distribution of \(\X_i\) is \(\text{Gumbel}(\theta_1, \theta_2, \delta)\). Based on the results of \cite{Zwetsloot_MainPaper}, we obtain the following conditional distributions used in the transformation (\ref{eqn:transform}):
\begin{equation}
\label{eqn:IC1_Gumbel}
F^{(0)}_{X_{(1)} \mid V}(x \mid v) = 1 - \exp\left(-[C(1,1)]^\delta x\right), \quad \text{for } v \in \{0,1,2\},
\end{equation}
\begin{equation}
\label{eqn:IC2_Gumbel}
F^{(0)}_{X_{(2)} \mid X_{(1)}, V}(y \mid x, v) = 
1 - \frac{\exp\left(-[C(x, y)]^\delta\right) [C(x, y)]^{\delta - 1}}{[C(x, x)]^{\delta - 1} \cdot \exp\left(-[C(1,1)]^\delta x\right)}, \quad \text{if } v = 0,
\end{equation}
\begin{equation}
\label{eqn:IC3_Gumbel}
F^{(0)}_{X_{(2)} \mid X_{(1)}, V}(y \mid x, v) = 
1 - \frac{\exp\left(-[C(y, x)]^\delta\right) [C(y, x)]^{\delta - 1}}{[C(x, x)]^{\delta - 1} \cdot \exp\left(-[C(1,1)]^\delta x\right)}, \quad \text{if } v = 1.
\end{equation}

Applying the transformation in (\ref{eqn:transform}) using the conditional distributions in (\ref{eqn:IC1_Gumbel})–(\ref{eqn:IC3_Gumbel}) results in transformed variables \(U_{i,(1)}\) and \(U_{i,(2)}\) that are i.i.d. \(\text{Unif}[0,1]\) under the IC process. The OC distribution of \(U_{i,(1)}\) is established in the following proposition.

\begin{proposition} \label{prop:OC_Gumbel}
Assume that \(\X_i\) follows an OC distribution \(\text{Gumbel}(\alpha_1, \alpha_2, \beta)\). Define \(D(x_1, x_2) = \left(x_1/\alpha_1\right)^{1/\beta} + \left(x_2/\alpha_2\right)^{1/\beta}\). Then the CDF of \(U_{i,(1)}\) is given by
\[
F_{U_{i,(1)}}(u) = 1 - (1 - u)^{[D(1,1)]^\beta / [C(1,1)]^\delta}.
\]
\end{proposition}

The above proposition establishes the OC distribution of \(U_{i,(1)}\), but obtaining a closed-form expression for the OC distribution of \(U_{i,(2)}\) is analytically intractable for general parameter values. However, in the special case where the components \(X_{i,1}\) and \(X_{i,2}\) are independent---namely when \(\delta = \beta = 1\)---a simplified expression is available. This result is stated in the following proposition.

\begin{proposition} \label{prop:OC2_Gumbel}
Assume that \(\X_i\) follows an OC distribution \(\text{Gumbel}(\alpha_1, \alpha_2, \beta)\). If \(\delta = \beta = 1\), then the transformed variables \(U_{i,(1)}\) and \(U_{i,(2)}\) are independent. Furthermore, the CDF of \(U_{i,(2)}\) is given by:
\[
F_{U_{i,(2)}}(u) = 
\begin{cases}
1 - (1 - u)^{\theta_2 / \alpha_2}, & \text{if } V_i = 0, \\
1 - (1 - u)^{\theta_1 / \alpha_1}, & \text{if } V_i = 1.
\end{cases}
\]
\end{proposition}

\subsection{Further Data Transformation}\label{sec:transform2}

Before proceeding further, it is worthwhile to reflect on the results from the previous subsection, as they are instrumental in motivating an additional transformation that simplifies the development of our final monitoring procedure.

Specifically, for all three distributions considered, the OC distribution of the first transformed variable \( U_{i,(1)} \) does not depend on \( V_i \). For the MOBE and MOBW cases (as well as the independent Gumbel case), the OC distribution of the second transformed variable \( U_{i,(2)} \) does depend on \( V_i \); however, it is independent of \( U_{i,(1)} \). This observation implies that there are only two possible distributions to consider for \( U_{i,(2)} \), significantly reducing the complexity of the modeling task. Moreover, the independence between \( U_{i,(1)} \) and \( U_{i,(2)} \) plays a key role in the construction of our monitoring statistic in the next section.

Importantly, across all OC distributions derived, the CDFs of both \( U_{i,(1)} \) and \( U_{i,(2)} \) share a unified functional form:
\[
F_U(u; k) = 1 - (1 - u)^k,
\]
where \( k > 0 \) is a constant that depends on both the IC and OC parameters. Notably, the IC distributions of \( U_{i,(1)} \) and \( U_{i,(2)} \) also follow the above form with \( k = 1 \).

Let \( U \sim F_U(u; k) \), and consider the transformation:
\[
Y = -\log(1 - U).
\]
Then the CDF of \( Y \) is given by
\[
F_Y(y; k) = \mathbb{P}(Y \leq y) = \mathbb{P}(-\log(1 - U) \leq y) = \mathbb{P}(U \geq 1 - e^{-y}) = 1 - e^{-k y},
\]
which corresponds to the exponential distribution with rate parameter \( k \), denoted by \( \text{Exp}(k) \).

This result is particularly useful. With the exception of \( U_{i,(2)} \) under the dependent Gumbel case, applying the transformation above to \( U_{i,(1)} \) and \( U_{i,(2)} \) yields new variables \( Y_{i,(1)} \) and \( Y_{i,(2)} \). Under the IC condition, both \( Y_{i,(1)} \) and \( Y_{i,(2)} \) follow \( \text{Exp}(1) \); under the OC condition, they follow \( \text{Exp}(k) \) with unknown rate \( k \).

This exponential representation provides a unified and tractable modeling framework that enables the construction of an efficient monitoring procedure based on exponential likelihoods. We present the details of this procedure in the following section.

\section{Adaptive CUSUM Chart for Bivariate TBE Data}

\subsection{Change-Point Model}\label{sec:CPM}
Recall that the original bivariate TBE observations, \( \X_1, \X_2, \ldots, \X_i, \ldots \), are observed sequentially through their order statistics \( X_{1,(1)}, X_{1,(2)}, X_{2,(1)}, X_{2,(2)}, \ldots, X_{i,(1)}, X_{i,(2)}, \ldots \). The indicator variable \( V_i \) defined in (\ref{eqn:Vi}) denotes which component of \( \X_i \) arrives first. As described in Section~\ref{sec:preliminary}, we apply the following transformation whenever $X_{i,(1)}$ or $X_{i,(2)}$ arrives, 
\[
Y_{i,(1)} = -\log\left(1 - F^{(0)}_{X_{(1)} \mid V}(X_{i,(1)} \mid V_i)\right), \quad 
Y_{i,(2)} = -\log\left(1 - F^{(0)}_{X_{(2)} \mid X_{(1)}, V}(X_{i,(2)} \mid X_{i,(1)}, V_i)\right),
\]
where \( F^{(0)}_{X_{(1)} \mid V} \) is the conditional CDF of \( X_{i,(1)} \) given \( V_i \), and \( F^{(0)}_{X_{(2)} \mid X_{(1)}, V} \) is the conditional CDF of \( X_{i,(2)} \) given \( X_{i,(1)} \) and \( V_i \), both under the IC distribution. Under the IC condition, the transformed sequence \( Y_{1,(1)}, Y_{1,(2)}, Y_{2,(1)}, Y_{2,(2)}, \ldots, Y_{i,(1)}, Y_{i,(2)}, \ldots \) consists of i.i.d.\ exponential random variables with rate 1.

From the results in Section~\ref{sec:preliminary}, under the OC condition and for all three distribution families considered, the variables \( Y_{i,(1)} \) remain independent and exponentially distributed with a common  rate parameter (denoted by \( k_1 \)), independent of \( V_i \). In contrast---except in the dependent Gumbel case---the OC distribution of \( Y_{i,(2)} \) is also exponential, but its rate parameter depends on \( V_i \). Specifically, when \( V_i = 0 \), \( Y_{i,(2)} \) follows an exponential distribution with a rate parameter denoted by \(k_2 \); when \( V_i = 1 \), \( Y_{i,(2)}\) follows another exponential distribution with a different rate parameter denoted by \(k_3 \).

Therefore, the change-point model of the original bivariate TBE data $\X_i$ in (\ref{eqn:CPM}) leads to the following change-point model for $Y_{i,(1)}$ and $Y_{i,(2)}$:
\[
Y_{i,(1)} \sim
\begin{cases}
   \text{Exp}(1), & \text{if } i \leq \tau \\
   \text{Exp}(k_1), & \text{if } i > \tau 
\end{cases} \quad \quad
Y_{i,(2)} \sim
\begin{cases}
   \text{Exp}(1), & \text{if } i \leq \tau \\
   \text{Exp}(k_2), & \text{if } i > \tau \text{ and } V_i = 0 \\
   \text{Exp}(k_3), & \text{if } i > \tau \text{ and } V_i = 1
\end{cases}
\]
Here, \( k_1, k_2, k_3 > 0 \) are unknown constants determined by both the IC and OC distributions of the original data \( \X_i \).

For notational simplicity, we relabel the transformed sequence \( Y_{1,(1)}, Y_{1,(2)}, Y_{2,(1)}, Y_{2,(2)}, \ldots\) as \( Z_1, Z_2, Z_3, Z_4,\ldots \). While each \( Z_t \) may follow one of three possible OC exponential distributions (i.e., \( \text{Exp}(k_1) \), \( \text{Exp}(k_2) \), or \( \text{Exp}(k_3) \)), we can determine exactly which one applies for each \( Z_t \). Specifically, define:
\begin{equation}
\label{eqn:Lt}
L_t = \begin{cases}
   1, & \text{if } Z_t = Y_{i,(1)} \text{ for some } i \\    
   2, & \text{if } Z_t = Y_{i,(2)} \text{ for some } i \text{ with } V_i = 0 \\
   3, & \text{if } Z_t = Y_{i,(2)} \text{ for some } i \text{ with } V_i = 1
\end{cases}
\end{equation}
Based on \(L_t\),  the OC distribution of \(Z_t\) can be written as Exp$(k_{L_t})$, where the value of \(L_t\) is fully determined once \(Z_t\) is observed. Consequently, the change-point model for \(Y_{i,(1)}\) and \(Y_{i,(2)}\) can be written more compactly as the following change-point model for \(Z_t\):
\[
Z_t \sim
\begin{cases}
   \text{Exp}(1), & \text{if } t \leq \tau^* \\
   \text{Exp}(k_{L_t}), & \text{if } t > \tau^*
\end{cases}
\]
where \( \tau^* \) is the change-point in the sequence \( \{Z_t\} \), and \( L_t \in \{1, 2, 3\} \) is known at time \( t \). This change-point model sets the stage for constructing a CUSUM chart that accounts for both the IC and OC distributions of \(Z_t\).

\subsection{CUSUM Statistic}
To develop a CUSUM chart based on the change-point model of \( Z_t \) described above, we begin with the general form of the CUSUM statistic originally introduced by \cite{Page1954}. This statistic is defined recursively as:
\[
C_t = \max\left(0, C_{t-1} + \log\left[\frac{f_1(Y_t)}{f_0(Y_t)}\right]\right),
\]
where \(C_0=0\), \( f_0 \) and \( f_1 \) are the probability density functions (PDFs) under the IC and OC distributions, respectively. A CUSUM chart monitors the sequence \( \{C_t\} \) over time and signals an alarm whenever \( C_t \) exceeds a predefined threshold \( h \), which is selected to satisfy a desired false alarm constraint.

The formulation above assumes a single OC distribution. However, in our setting, each observation \( Z_t \) may follow one of three distinct OC exponential distributions depending on the value of \( L_t \in \{1, 2, 3\} \). Therefore, this standard CUSUM formulation does not directly apply. Since the above CUSUM statistic is derived from the log-likelihood ratio test, we can extend it to accommodate multiple OC distributions by using the same likelihood ratio principle. In our case, the CUSUM statistic is generalized to:
\[
C_t = \max\left(0, C_{t-1} + \log\left[\frac{f_{L_t}(Z_t)}{f_0(Z_t)}\right]\right),
\]
where \(C_0=0\), \( f_0 \) is the PDF of the IC distribution \( \text{Exp}(1) \), and \( f_{L_t} \) is the PDF of the corresponding OC distribution \( \text{Exp}(k_{L_t}) \), as determined by the observed value of \( L_t \). Substituting the PDFs of these exponential distributions into the expression yields the final form of our CUSUM statistic:
\begin{equation}
\label{eqn:CUSUM_3exp}
C_t = \max\left(0, C_{t-1} + \log(k_{L_t}) + (1 - k_{L_t}) Z_t \right).
\end{equation}

\subsection{Adaptive CUSUM Statistics}\label{sec:AdaptiveCUSUM}

The CUSUM statistic defined in~\eqref{eqn:CUSUM_3exp} requires the specification of the OC parameters \( k_1 \), \( k_2 \), and \( k_3 \). However, these parameters are typically unknown in practice. To address this, we adopt the \textit{adaptive CUSUM} framework pioneered by \cite{Lorden2008}, which preserves the recursive structure of the CUSUM statistic while adaptively estimating the OC parameters. This approach is both computationally efficient and asymptotically optimal up to second order for single-parameter exponential families. \cite{Wu2017} further extended the method to multi-parameter exponential families, establishing its first-order asymptotic optimality.

The key idea behind an adaptive CUSUM procedure is to mimic the CUSUM statistic in~\eqref{eqn:CUSUM_3exp}, but replace the unknown OC parameter \( k_{L_t} \) with an adaptive estimate \( \widehat{k}_{L_t,t} \). Specifically, the adaptive CUSUM statistic is defined recursively as:
\begin{equation*}
\widehat{C}_t = \max\left(0, \widehat{C}_{t-1} + \log(\widehat{k}_{L_t,t}) + (1 - \widehat{k}_{L_t,t})Z_t\right),
\end{equation*}
with \(\widehat{C}_{0}=0\).

We now explain how to construct the estimate \( \widehat{k}_{L_t,t} \) at each time \( t \). Ideally, if the change-point \( \tau^* \) were known, one could estimate \( k_{L_t} \) using only the post-change observations. However, since \( \tau^* \) is unknown, we follow the approach of \cite{Lorden2008} and \cite{Wu_ChangepointEstim_2005}, and define the estimated change-point at time \( t \), denoted by \( \widehat{\tau}_t \), as the most recent time at which the adaptive CUSUM statistic \( \widehat{C}_t \) resets to zero. We then estimate \( k_{L_t} \) using the observations \( Z_s \) in the interval \( [\widehat{\tau}_t+1, t-1] \) that follow the same OC distribution \( \text{Exp}(k_{L_t}) \). The current observation \( Z_t \) is excluded from the estimation to avoid ``double dipping'' and preserve the recursive and martingale properties of \( \widehat{C}_t \) under the IC hypothesis.

Let \( N_{L_t,t} \) and \( S_{L_t,t} \) denote the count and sum, respectively, of observations \( Z_s\)  with OC distribution \(\text{Exp}(k_{L_t}) \) in the interval \( [\widehat{\tau}_t+1, t-1] \). A natural maximum likelihood estimator (MLE) of \( k_{L_t} \) is then:
\[
\widehat{k}_{L_t,t} = \frac{N_{L_t,t}}{S_{L_t,t}}.
\]
However, when \( t \) is close to \( \widehat{\tau}_t \), the sample size is small and the estimator can be highly variable. To stabilize the estimation, we adopt the Bayesian approach of \cite{Liu_etal2018} and place a conjugate Gamma prior, \( \text{Gamma}(\alpha, \beta) \), on \( k_{L_t} \). The resulting Bayesian estimator is:
\[
\widehat{k}_{L_t,t} = \frac{\alpha + N_{L_t,t}}{\beta + S_{L_t,t}}.
\]
To tailor the estimator for detecting increases or decreases in \( k_{L_t} \), we use different priors. For detecting increases, we set the prior expectation of \( k_{L_t} \) to 1.05 by choosing \( \alpha^+ = 22.05 \) and \( \beta^+ = 21 \). For detecting decreases, we set the prior expectation of \( k_{L_t} \) to 0.95 using \( \alpha^- = 9.5 \) and \( \beta^- = 10 \).

Because \( L_t \in \{1, 2, 3\} \), we construct two Bayesian estimators (for increase and decrease) for each \( k_{L_t} \), resulting in \( 2^3 = 8 \) combinations. We denote each corresponding adaptive CUSUM statistic as:
\[
\widehat{C}_t^{(*_1, *_2, *_3)}, \quad \text{where } (*_1, *_2, *_3) \in \{+, -\}^3,
\]
and \( *_j = + \) or \( - \) indicates whether the CUSUM is designed to detect an increase or a decrease in \( k_j \). For example, \( \widehat{C}_t^{(+, -, +)} \) targets  the OC scenario in which \( k_1 \) and \( k_3 \) increase while \( k_2 \) decreases.

We now outline the recursive algorithm used to compute each adaptive CUSUM statistic \(\widehat{C}_t^{(*_1, *_2, *_3)}\), where \( (*_1, *_2, *_3) \in \{+, -\}^3 \). Let \( \widehat{\tau}_t^{(*_1, *_2, *_3)} \) denote the most recent time at which \( \widehat{C}_t^{(*_1, *_2, *_3)} \) resets to zero. For $j \in \{1,2,3\}$, define:
\[
N_{j,t}^{(*_1, *_2, *_3)} = \sum_{s = \widehat{\tau}_t^{(*_1, *_2, *_3)}+1}^{t-1} I(L_s = j), \quad
S_{j,t}^{(*_1, *_2, *_3)} = \sum_{s = \widehat{\tau}_t^{(*_1, *_2, *_3)}+1}^{t-1} Z_s \cdot I(L_s = j),
\]
as the count and sum, respectively, of observations \( Z_s \) with OC distribution 
\( \text{Exp}(k_{j}) \)
in the interval \( [\widehat{\tau}^{(*_1, *_2, *_3)}_t+1, t-1] \). These quantities can be updated recursively via:
\[
N_{j,t}^{(*_1, *_2, *_3)} =
\begin{cases}
   N_{j,t-1}^{(*_1, *_2, *_3)} + 1 & \text{if } \widehat{C}_{t-1}^{(*_1, *_2, *_3)} > 0 \text{ and } L_{t-1} = j,\\    
   0 & \text{if } \widehat{C}_{t-1}^{(*_1, *_2, *_3)} = 0 \text{ and } L_{t-1} = j,\\
   N_{j,t-1}^{(*_1, *_2, *_3)} & \text{if } L_{t-1} \neq j,
\end{cases}
\]
\[
S_{j,t}^{(*_1, *_2, *_3)} =
\begin{cases}
   S_{j,t-1}^{(*_1, *_2, *_3)} + Z_{t-1} & \text{if } \widehat{C}_{t-1}^{(*_1, *_2, *_3)} > 0 \text{ and } L_{t-1} = j,\\    
   0 & \text{if } \widehat{C}_{t-1}^{(*_1, *_2, *_3)} = 0 \text{ and } L_{t-1} =j,\\
   S_{j,t-1}^{(*_1, *_2, *_3)} & \text{if } L_{t-1} \neq j,
\end{cases}
\]
with \(N_{1,0}^{(*_1, *_2, *_3)}=N_{2,0}^{(*_1, *_2, *_3)}=N_{3,0}^{(*_1, *_2, *_3)}=S_{1,0}^{(*_1, *_2, *_3)}=S_{2,0}^{(*_1, *_2, *_3)}=S_{3,0}^{(*_1, *_2, *_3)}=0\).
The Bayesian estimator for \( k_{j} \) is then given by:
\[
\widehat{k}_{j,t}^{(*_1, *_2, *_3)} =
\begin{cases}
   \max\left(\rho^+, \dfrac{ \alpha^+ + N_{j,t}^{(*_1, *_2, *_3)} }{\beta^+ + S_{j,t}^{(*_1, *_2, *_3)}} \right) & \text{if } *_{j} = +,\\    
   \min\left(\rho^-, \dfrac{ \alpha^- + N_{j,t}^{(*_1, *_2, *_3)} }{\beta^- + S_{j,t}^{(*_1, *_2, *_3)}} \right) & \text{if } *_{j} = -,
\end{cases}
\]
where \( \rho^+ = 1.05 \) and \( \rho^- = 0.95 \) serve as minimal meaningful shifts to bound the estimator away from 1.

Substituting \(\widehat{k}_{L_t,t}^{(*_1, *_2, *_3)} \) into~\eqref{eqn:CUSUM_3exp}, we obtain the final adaptive CUSUM statistic:
\begin{equation}
\label{eqn:adaptiveCUSUM_3exp}
\widehat{C}_{t}^{(*_1, *_2, *_3)} = \max\left(0, \widehat{C}_{t-1}^{(*_1, *_2, *_3)} + \log(\widehat{k}_{L_t,t}^{(*_1, *_2, *_3)}) + (1 - \widehat{k}_{L_t,t}^{(*_1, *_2, *_3)}) Z_t \right),
\end{equation}
with \( \widehat{C}_{0}^{(*_1, *_2, *_3)}=0 \)  for all \( (*_1, *_2, *_3) \in \{+, -\}^3 \).

\subsection{Aggregation of Adaptive CUSUM Statistics}

Based on the two choices of prior parameters in Section~\ref{sec:AdaptiveCUSUM}, each of the eight adaptive CUSUM statistics \( \widehat{C}^{(*_1, *_2, *_3)}_{t} \), where \( (*_1, *_2, *_3) \in \{+, -\}^3 \), defined in~\eqref{eqn:adaptiveCUSUM_3exp}, is more powerful for detecting a specific OC scenario involving increases or decreases in \( (k_1, k_2, k_3) \). In the absence of prior knowledge about the type of change the process may undergo, it is necessary to monitor all eight adaptive CUSUM statistics simultaneously and combine them into a single charting statistic. If these adaptive CUSUM statistics shared a common IC distribution, aggregation could be performed via summation or by taking the maximum. However, in our case, the eight statistics do not share a common IC distribution, so direct comparison or aggregation is not feasible.

To circumvent this issue, we can find the IC distribution of each adaptive CUSUM statistic and then apply the probability integral transform to map them to a common distribution. Since an analytical derivation of the IC distribution is infeasible, we resort to empirical approximation at each time point. At first glance, this task appears daunting: the distribution of each statistic evolves over time and the time index \( t \) is unbounded. However, this challenge can be partially mitigated by noting that the IC evolution of each adaptive CUSUM statistic can be modeled as a Markov chain. In particular, the chain is irreducible, aperiodic, and positively recurrent, ensuring convergence to a unique stationary distribution. Therefore, it suffices to approximate the distribution during the early stages of evolution; beyond a certain point, the statistic can be treated as drawn from its stationary distribution.


Define the IC CDF of the adaptive CUSUM statistic \( \widehat{C}_{t}^{(*_{1}, *_{2}, *_{3})} \), for \( (*_1, *_2, *_3) \in \{+,-\}^3 \), conditional on the statistic being nonzero, as
\[
F_t^{(*_{1}, *_{2}, *_{3})}(c) := P^{(0)}\left(\widehat{C}_{t}^{(*_{1}, *_{2}, *_{3})} < c \,\middle|\, \widehat{C}_{t}^{(*_{1}, *_{2}, *_{3})} \neq 0\right).
\]
Suppose that we have a good approximation of \( F_t^{(*_{1}, *_{2}, *_{3})} \), denoted by \( \widehat{F}_t^{(*_{1}, *_{2}, *_{3})} \), for all \( t \). Define the transformed statistic as
\[
p^{(*_{1}, *_{2}, *_{3})}_t = \widehat{F}_t^{(*_{1}, *_{2}, *_{3})}\left(\widehat{C}_{t}^{(*_{1}, *_{2}, *_{3})} \right).
\]
By the probability integral transform, each \( p^{(*_{1}, *_{2}, *_{3})}_t \) follows a \text{Unif}(0,1)  distribution under the IC condition when the corresponding adaptive CUSUM statistic is nonzero. Since these transformed statistics share a common IC distribution, they can be aggregated as $
p_{t} = \max_{(*_1, *_2, *_3) \in \{+,-\}^3 } p_{t}^{(*_1, *_2, *_3)}$.
Although one could monitor \( p_{t} \) directly to detect OC behavior, we instead apply a transformation to each \(p_{t}^{(*_1, *_2, *_3)}\) to obtain a positive, unbounded statistic, as this can simplify the estimation of control limits. Specifically, we define:
\[
Q_{t}^{(*_1, *_2, *_3)}=-\log(1-p_{t}^{(*_1, *_2, *_3)}).
\]
By the inverse CDF transform, each \( Q_{t}^{(*_1, *_2, *_3)} \) follows Exp(1) under the IC condition when \( \widehat{C}_{t}^{(*_{1}, *_{2}, *_{3})} \neq 0 \). Accordingly, we aggregate these statistics as:
\[
Q_{t} = \max_{(*_1, *_2, *_3) \in \{+,-\}^3} Q_{t}^{(*_{1}, *_{2}, *_{3})}.
\]
Our final adaptive CUSUM chart monitors the sequence \( Q_t \) over time and raises an alarm if \( Q_t \) exceeds a predefined threshold \( h \). This threshold \( h \) is selected to satisfy a desired false alarm constraint and can be determined using a numerical method such as the bisection algorithm.

\section{Simulation Studies}\label{sec:simulation}

In this section, we conduct simulation studies to evaluate the performance of our proposed adaptive CUSUM chart for real-time monitoring of bivariate TBE data. We compare it against the Shewhart chart developed by \cite{Zwetsloot_MainPaper}, replicating their simulation settings exactly. Specifically, we assess both methods across the three bivariate distributions discussed in Section~\ref{sec:OCdist}: the MOBE, MOBW, and Gumbel distributions. The parameter values for each distribution are taken directly from Table B4 in \cite{Zwetsloot_MainPaper}. For each distribution, the following four IC scenarios are considered:
\begin{itemize}
    \item \textbf{Scenario 1:} \( E[X_1] = E[X_2] = 5 \), with \( X_1 \) and \( X_2 \) independent.
    \item \textbf{Scenario 2:} \( E[X_1] = E[X_2] = 5 \), with \( X_1 \) and \( X_2 \) dependent.
    \item \textbf{Scenario 3:} \( E[X_1] = 5 \), \( E[X_2] = 15 \), with \( X_1 \) and \( X_2 \) independent.
    \item \textbf{Scenario 4:} \( E[X_1] = 5 \), \( E[X_2] = 15 \), with \( X_1 \) and \( X_2 \) dependent.
\end{itemize}

To simplify implementation, instead of empirically estimating the time-varying IC distributions of the adaptive CUSUM statistics until they stabilize, we initialize each statistic and its associated estimators using samples drawn directly from their respective stationary (steady-state) distributions. This approach ensures that all subsequent statistics remain within the stationary regime, eliminating the need to store the full time-varying distributions. Moreover, starting in steady state yields steady-state performance results, which are commonly used in control chart evaluations.

Performance comparisons between the proposed adaptive CUSUM chart and the Shewhart chart of \cite{Zwetsloot_MainPaper} under different distribution settings are reported in Tables~\ref{tab:MOBE}--\ref{tab:Gumbel}. In each table, the first row of every scenario corresponds to the IC case, while the remaining eight rows represent various OC shifts, including increases or decreases in one or both components.

To maintain consistency with \cite{Zwetsloot_MainPaper}, we use the average time to signal (ATS) as our primary performance metric. The ATS measures the expected elapsed time until a signal is raised and is equivalent to the more commonly used average run length (ARL) metric, in the sense that there exists a one-to-one correspondence between the two. For both monitoring procedures, control limits are chosen such that the IC ATS is 200. Tables~\ref{tab:MOBE}--\ref{tab:Gumbel} report the ATS values for each IC and OC setting, with standard errors shown in parentheses. Under OC conditions, lower ATS values indicate faster detection and thus better performance.
\begin{table}[!htpb]
\centering
\scalebox{1.2}{
\def\arraystretch{.6}
\begin{tabular}{llll}
\toprule
Scenario & Mean & Adaptive CUSUM & Shewhart \\
\midrule
1 & ( 5 , 5 ) & 199.3 (2.95) & 198.5 (1.94) \\
1 & ( 7.5 , 5 ) & 109.5 (1.34) & 142.8 (1.28) \\
1 & ( 10 , 5 ) & 72.9 (0.782) & 103.4 (0.847) \\
1 & ( 7.5 , 7.5 ) & 79.2 (0.87) & 108.3 (0.928) \\
1 & ( 10 , 10 ) & 53.6 (0.51) & 73.4 (0.571) \\
1 & ( 5 , 2.5 ) & 60.8 (0.7) & 171.4 (1.7) \\
1 & ( 5 , 1 ) & 27 (0.268) & 100.7 (1.02) \\
1 & ( 2.5 , 2.5 ) & 23 (0.239) & 99.1 (1.01) \\
1 & ( 1 , 1 ) & 5.4 (0.0501) & 16 (0.169) \\
\addlinespace
2 & ( 5 , 5 ) & 196.7 (2.98) & 203.5 (1.98) \\
2 & ( 7.5 , 5 ) & 110.9 (1.38) & 145.2 (1.33) \\
2 & ( 10 , 5 ) & 72.9 (0.776) & 107.1 (0.883) \\
2 & ( 7.5 , 7.5 ) & 76.3 (0.852) & 112 (0.967) \\
2 & ( 10.1 , 10.1 ) & 52.9 (0.49) & 73.1 (0.562) \\
2 & ( 5 , 2.5 ) & 61.8 (0.745) & 176.9 (1.73) \\
2 & ( 1 , 5 ) & 26.4 (0.272) & 107.6 (1.05) \\
2 & ( 2.5 , 2.5 ) & 23.6 (0.246) & 102.6 (1.05) \\
2 & ( 1 , 1 ) & 5.5 (0.0514) & 16.5 (0.172) \\
\addlinespace
3 & ( 5 , 15 ) & 193.6 (3.37) & 194.1 (1.9) \\
3 & ( 7.5 , 15 ) & 110.2 (1.7) & 135.2 (1.26) \\
3 & ( 10 , 15 ) & 72.8 (0.981) & 99.2 (0.87) \\
3 & ( 7.5 , 22.7 ) & 97.7 (1.34) & 132.4 (1.14) \\
3 & ( 10 , 30.3 ) & 73.9 (0.886) & 102.1 (0.805) \\
3 & ( 5 , 10.5 ) & 107.6 (1.75) & 167.4 (1.67) \\
3 & ( 5 , 7.5 ) & 59.5 (0.845) & 128.1 (1.27) \\
3 & ( 3.5 , 7.5 ) & 41.8 (0.568) & 119.1 (1.24) \\
3 & ( 2.5 , 7.5 ) & 31.1 (0.412) & 99.8 (1.04) \\
\addlinespace
4 & ( 5 , 15 ) & 206 (3.58) & 203.2 (1.98) \\
4 & ( 7.5 , 15 ) & 114.8 (1.77) & 140.4 (1.29) \\
4 & ( 10 , 15 ) & 78.1 (1.06) & 103.9 (0.918) \\
4 & ( 7.5 , 22.7 ) & 104.6 (1.43) & 140.1 (1.2) \\
4 & ( 10 , 30.3 ) & 79.4 (0.941) & 104.7 (0.827) \\
4 & ( 5 , 10.5 ) & 109.1 (1.74) & 172 (1.72) \\
4 & ( 5 , 7.5 ) & 61.1 (0.843) & 130.5 (1.34) \\
4 & ( 3.5 , 7.5 ) & 42.1 (0.566) & 120.3 (1.24) \\
4 & ( 2.5 , 7.5 ) & 31.7 (0.414) & 99.4 (1.06) \\
\bottomrule
\end{tabular}}
\caption{Performance comparison between our proposed adaptive CUSUM chart and the Shewhart chart under the MOBE distribution.\label{tab:MOBE}}
\end{table}

\begin{table}[!htpb] 
\centering
\scalebox{1.2}{
\def\arraystretch{.6}
\begin{tabular}{llll}
\toprule
Scenario & Mean & Adaptive CUSUM & Shewhart \\
\midrule
1 & ( 5 , 5 ) & 201.4 (2.92) & 203.2 (2.01) \\
1 & ( 7.5 , 5 ) & 47 (0.471) & 68 (0.59) \\
1 & ( 10 , 5 ) & 29.9 (0.244) & 36 (0.252) \\
1 & ( 7.5 , 7.5 ) & 33 (0.285) & 41.4 (0.327) \\
1 & ( 10 , 10 ) & 23.4 (0.16) & 22.8 (0.134) \\
1 & ( 5 , 2.5 ) & 34.7 (0.308) & 133 (1.32) \\
1 & ( 5 , 1 ) & 19.6 (0.166) & 37.8 (0.4) \\
1 & ( 2.5 , 2.5 ) & 14.2 (0.111) & 50.2 (0.518) \\
1 & ( 1 , 1 ) & 4.4 (0.0323) & 3.5 (0.0366) \\
\addlinespace
2 & ( 5 , 5 ) & 198.6 (2.99) & 199 (1.98) \\
2 & ( 7.5 , 5 ) & 47.4 (0.477) & 66.8 (0.574) \\
2 & ( 10 , 5 ) & 29.9 (0.244) & 35.6 (0.247) \\
2 & ( 7.5 , 7.5 ) & 33.4 (0.286) & 41.4 (0.323) \\
2 & ( 10 , 10 ) & 23.6 (0.162) & 22.9 (0.134) \\
2 & ( 5 , 2.5 ) & 33.8 (0.31) & 135.9 (1.39) \\
2 & ( 5 , 2 ) & 27.1 (0.232) & 107.4 (1.08) \\
2 & ( 2.5 , 2.5 ) & 14.2 (0.112) & 49 (0.498) \\
2 & ( 1 , 1 ) & 4.4 (0.0327) & 3.4 (0.036) \\
\addlinespace
3 & ( 5 , 15 ) & 206.9 (3.42) & 195.5 (1.9) \\
3 & ( 7.5 , 15 ) & 53 (0.673) & 72 (0.67) \\
3 & ( 10 , 15 ) & 32.9 (0.333) & 39.1 (0.323) \\
3 & ( 7.5 , 22.2 ) & 49.5 (0.549) & 63.2 (0.516) \\
3 & ( 10 , 29.5 ) & 36.8 (0.332) & 39 (0.284) \\
3 & ( 5 , 10.5 ) & 64.9 (0.819) & 150.8 (1.54) \\
3 & ( 5 , 7.5 ) & 37 (0.397) & 88 (0.873) \\
3 & ( 3.5 , 7.5 ) & 25.3 (0.26) & 69.6 (0.724) \\
3 & ( 2.5 , 7.5 ) & 20.1 (0.202) & 49.8 (0.528) \\
\addlinespace
4 & ( 5 , 15 ) & 200.4 (3.27) & 200.9 (1.94) \\
4 & ( 7.5 , 15 ) & 53 (0.674) & 75.7 (0.707) \\
4 & ( 10 , 15 ) & 33.4 (0.331) & 39.6 (0.323) \\
4 & ( 7.5 , 22.6 ) & 49.7 (0.53) & 64.3 (0.517) \\
4 & ( 10 , 29.9 ) & 37.4 (0.327) & 39.7 (0.275) \\
4 & ( 5 , 10.5 ) & 65.6 (0.806) & 160.8 (1.62) \\
4 & ( 5 , 7.5 ) & 36.3 (0.392) & 91.9 (0.919) \\
4 & ( 3.5 , 7.5 ) & 25.4 (0.256) & 72.8 (0.735) \\
4 & ( 2.5 , 7.5 ) & 19.9 (0.198) & 52.3 (0.552) \\
\bottomrule
\end{tabular}}
\caption{Performance comparison between our proposed adaptive CUSUM chart and the Shewhart chart under the MOBW distribution.\label{tab:MOBW}}
\end{table}

\begin{table}[!htpb]
\centering
\scalebox{1.2}{
\def\arraystretch{.6}
\begin{tabular}{llll}
\toprule
Scenario & Mean & Adaptive CUSUM & Shewhart \\
\midrule
1 & ( 5 , 5 ) & 200.2 (3.02) & 198.8 (1.94) \\
1 & ( 7.5 , 5 ) & 109.2 (1.35) & 141.8 (1.28) \\
1 & ( 10 , 5 ) & 71.6 (0.751) & 104.8 (0.854) \\
1 & ( 7.5 , 7.5 ) & 78.1 (0.854) & 109.3 (0.945) \\
1 & ( 10 , 10 ) & 54.1 (0.516) & 73.1 (0.571) \\
1 & ( 5 , 2.5 ) & 60.8 (0.701) & 172.3 (1.7) \\
1 & ( 5 , 1 ) & 26.8 (0.269) & 101.1 (1.02) \\
1 & ( 2.5 , 2.5 ) & 23.2 (0.239) & 100.1 (1) \\
1 & ( 1 , 1 ) & 5.6 (0.0511) & 16.1 (0.167) \\
\addlinespace
2 & ( 5 , 5 ) & 201.3 (2.94) & 198.7 (1.93) \\
2 & ( 7.5 , 5 ) & 109 (1.28) & 149.6 (1.36) \\
2 & ( 10 , 5 ) & 68.1 (0.654) & 108.3 (0.902) \\
2 & ( 7.5 , 7.5 ) & 90.2 (1.01) & 126.3 (1.13) \\
2 & ( 10 , 10 ) & 62.7 (0.576) & 85.8 (0.69) \\
2 & ( 5 , 2.5 ) & 51.8 (0.564) & 163 (1.61) \\
2 & ( 5 , 1 ) & 17.2 (0.152) & 66.6 (0.602) \\
2 & ( 2.5 , 2.5 ) & 26.5 (0.273) & 117.6 (1.19) \\
2 & ( 1 , 1 ) & 6.1 (0.055) & 24.8 (0.259) \\
\addlinespace
3 & ( 5 , 15 ) & 189.7 (3.33) & 196 (1.9) \\
3 & ( 7.5 , 15 ) & 109.1 (1.73) & 137.8 (1.29) \\
3 & ( 10 , 15 ) & 72.2 (1) & 100 (0.898) \\
3 & ( 7.5 , 22.5 ) & 100.8 (1.39) & 134.6 (1.16) \\
3 & ( 10 , 30 ) & 75.7 (0.914) & 100.4 (0.805) \\
3 & ( 5 , 10.5 ) & 107.4 (1.74) & 167 (1.7) \\
3 & ( 5 , 7.5 ) & 59.3 (0.824) & 126.4 (1.28) \\
3 & ( 3.5 , 7.5 ) & 41.4 (0.567) & 118.5 (1.23) \\
3 & ( 2.5 , 7.5 ) & 31.9 (0.42) & 98.9 (1.06) \\
\addlinespace
4 & ( 5 , 15 ) & 204.4 (3.47) & 197.6 (1.94) \\
4 & ( 7.5 , 15 ) & 91.8 (1.28) & 124.3 (1.15) \\
4 & ( 10 , 15 ) & 64 (0.752) & 86.7 (0.745) \\
4 & ( 7.5 , 22.5 ) & 118.4 (1.65) & 144.2 (1.27) \\
4 & ( 10 , 30 ) & 89.2 (1.05) & 111.7 (0.925) \\
4 & ( 5 , 10.5 ) & 89.5 (1.34) & 147.8 (1.49) \\
4 & ( 5 , 7.5 ) & 47.9 (0.625) & 96.4 (0.956) \\
4 & ( 3.5 , 7.5 ) & 44.5 (0.612) & 111.1 (1.13) \\
4 & ( 2.5 , 7.5 ) & 37.3 (0.505) & 116 (1.2) \\
\bottomrule
\end{tabular}}
\caption{Performance comparison between our proposed adaptive CUSUM chart and the Shewhart chart under the Gumbel distribution.\label{tab:Gumbel}}
\end{table}

As seen in Tables~\ref{tab:MOBE}-\ref{tab:Gumbel}, our proposed adaptive CUSUM chart maintains good control of the IC ATS and achieves lower ATS values than the Shewhart chart in nearly all OC settings. Although the adaptive CUSUM chart consistently outperforms the Shewhart chart in both mean-increase and mean-decrease scenarios, the performance gap is especially pronounced when the mean decreases. (In many real-world applications, decreases in the mean of TBE are typically more concerning than increases.) While it is well known that detection delay decreases as the size of the mean shift increases, it is often overlooked that the sensitivity of Shewhart-type charts also depends heavily on the variability of the observations. For example, when the OC distribution has a smaller variance than the IC distribution, the Shewhart chart may struggle to generate timely signals, since observations are less likely to exceed the control limits due to the reduced variability. This phenomenon helps explain the substantial detection delays observed in the Shewhart chart when the OC mean decreases, since the OC variance is also reduced. In contrast, our adaptive CUSUM chart accounts for changes in both the mean and variance, as it directly incorporates the OC distribution in its construction. Furthermore, the ability of our adaptive CUSUM chart to accumulate evidence over time also makes it more efficient in detecting small to moderate shifts.

It is worth noting that in the dependent Gumbel case, the transformed variables under the OC regime may no longer be independent or exponentially distributed. This implies that the OC distributions used in constructing our adaptive CUSUM chart may be misspecified in this setting. Nevertheless, as shown in Table~\ref{tab:Gumbel}, our chart still significantly outperforms the Shewhart chart, indicating that the proposed procedure is robust to moderate levels of model misspecification.

\section{Real Data Application}

To evaluate the practical utility of our proposed method, we apply it to a real-world dataset previously studied by \cite{Zwetsloot_MainPaper}. This dataset originates from a collaboration between the City University of Hong Kong and a local company responsible for maintaining many of the city's escalators (\cite{HU2024}).

As described in \cite{Zwetsloot_MainPaper}, escalators are complex mechanical systems with a typical lifespan of around 30 years and are expected to operate reliably under heavy daily use. Faults may occur due to internal component failures or external disruptions such as passenger misuse or foreign object interference. To monitor escalator health, \cite{Zwetsloot_MainPaper} proposed two key indicators: the \textit{time between faults} ($X_1$) and the \textit{time to repair} each fault ($X_2$). In general, longer fault-free intervals and shorter repair times reflect better operational performance and maintenance efficiency.

Because a fault must occur before it can be repaired, $X_1$ is always observed prior to $X_2$. However, to allow for modeling scenarios where either event could occur first—common in bivariate failure models—\cite{Zwetsloot_MainPaper} recorded $X_1$ in days and $X_2$ in minutes. This unit transformation, though somewhat artificial, enables more flexible statistical modeling and reflects the asynchronous nature of typical bivariate TBE data. Therefore, the dataset serves as a valuable test case for evaluating our adaptive CUSUM chart under realistic operational conditions with inherent dependencies between components.

The dataset consists of 245 pairs of fault-repair event times. Following \cite{Zwetsloot_MainPaper}, we treat the first 100 pairs as Phase I (IC) data and the remaining 145 as Phase II (monitoring) data. Using the \texttt{fitdistrplus} package in R, \cite{Zwetsloot_MainPaper} fitted a MOBW distribution to the Phase I data and obtained the following IC parameter estimates:
\[
\eta = 1.1677, \quad \lambda_1 = 0.0435, \quad \lambda_2 = 0.0105, \quad \lambda_3 = 5.78 \times 10^{-8}.
\]
To ensure a fair comparison, we adopt the same IC distribution for our proposed adaptive CUSUM chart. Consistent with their setup, we calibrate the control limits to achieve an IC ARL (denoted by $\text{ARL}_0 $) of 50.

We then apply our monitoring procedure to the Phase II observations. Our adaptive CUSUM chart signals an alarm at the 9\textsuperscript{th} Phase II observation (i.e., the first observation from the 5\textsuperscript{th} vector), whereas the Shewhart chart proposed in \cite{Zwetsloot_MainPaper} signals at the 92\textsuperscript{nd} Phase II observation (i.e., the second observation from the 46\textsuperscript{th} vector).

\begin{figure}[!htpb]
\centering
\includegraphics[width=.8\textwidth]{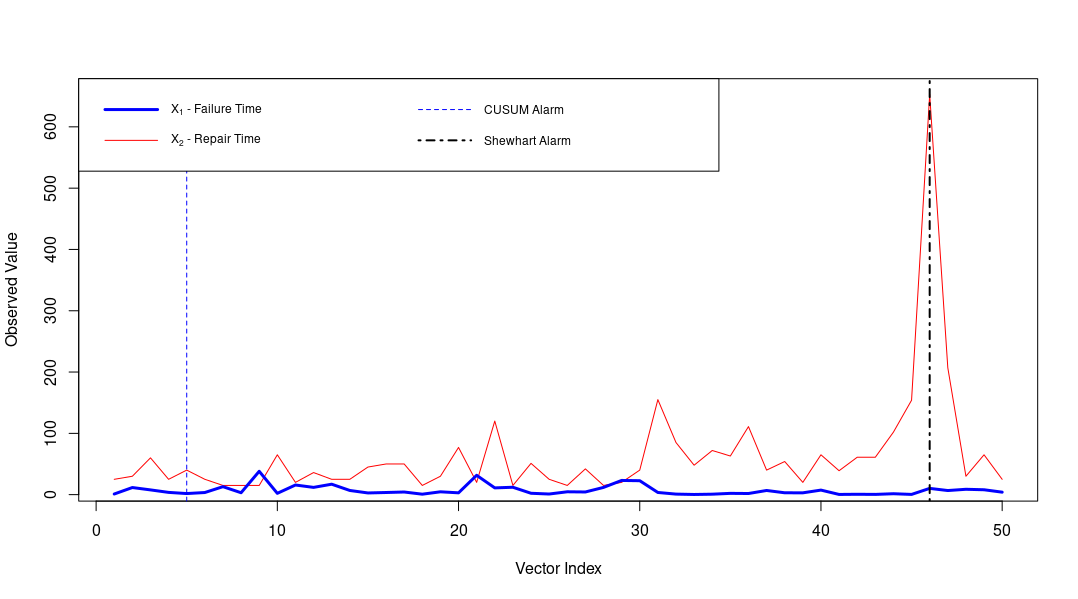}
\caption{Visualization of the first 50 Phase II vectors along with the alarm times from our proposed adaptive CUSUM chart and the Shewhart chart.}
\label{fig:rawdata}
\end{figure}

To visualize this difference, we plot the first 50 Phase II vectors (corresponding to the first 100 observations), with alarm times indicated for both our adaptive CUSUM chart and the Shewhart chart, as shown in Figure~\ref{fig:rawdata}.
For the Shewhart chart, the alarm is easily explained by a conspicuously large repair time. However, the triggering mechanism for our adaptive CUSUM chart is less visually apparent in the raw observation space.

\begin{figure}[!htpb]
\centering
\includegraphics[width=.9\textwidth]{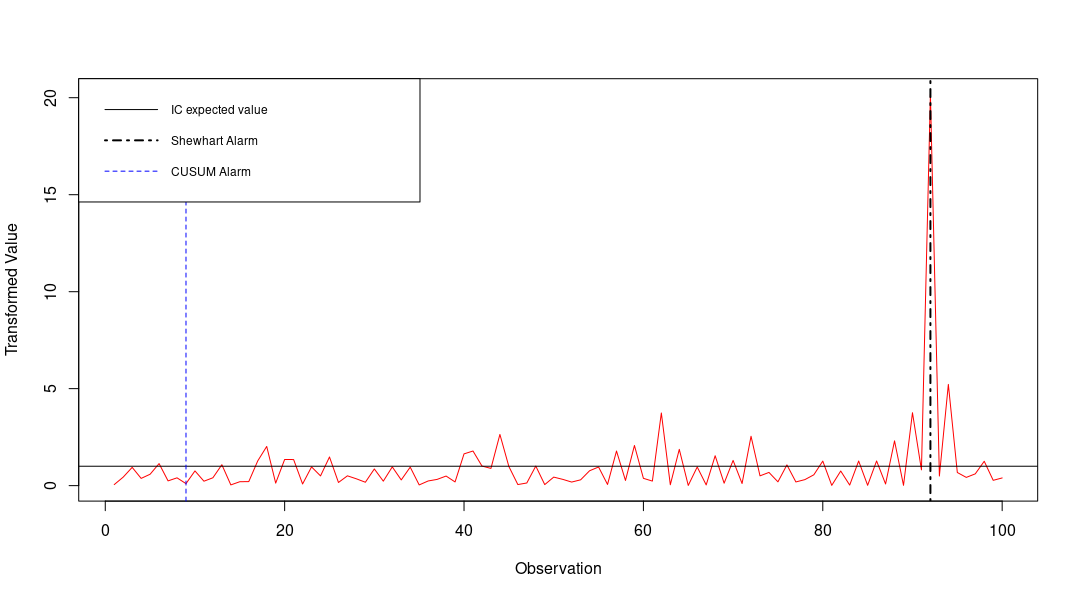}
\caption{The first 100 transformed observations $Z_t$, with the corresponding CUSUM and Shewhart alarms. The solid black line marks the expected value of $Z_t$ under the IC distribution.}
\label{fig:transformeddata}
\end{figure}

Recall that in our proposed adaptive CUSUM procedure, we first transform the raw, bivariate, correlated data $\X_i$ into independent, exponentially distributed univariate data $Z_t$ before applying the adaptive CUSUM chart. Consequently, it is more informative to examine the $Z_t$'s in the transformed space. Under the IC model, the $Z_t$'s should be i.i.d.\ $\text{Exp}(1)$ random variables. Figure~\ref{fig:transformeddata} displays the first 100 transformed observations $Z_t$, with the solid black line indicating the expected value under the IC distribution.

In the transformed space, we again observe that the Shewhart alarm is triggered by a single large outlier—consistent with an unusually long repair time. In contrast, the CUSUM alarm is driven by a sequence of observations that systematically fall below the expected value, suggesting a sustained downward shift in the process mean.

\begin{figure}[!htpb]
\centering
\includegraphics[width=.9\textwidth]{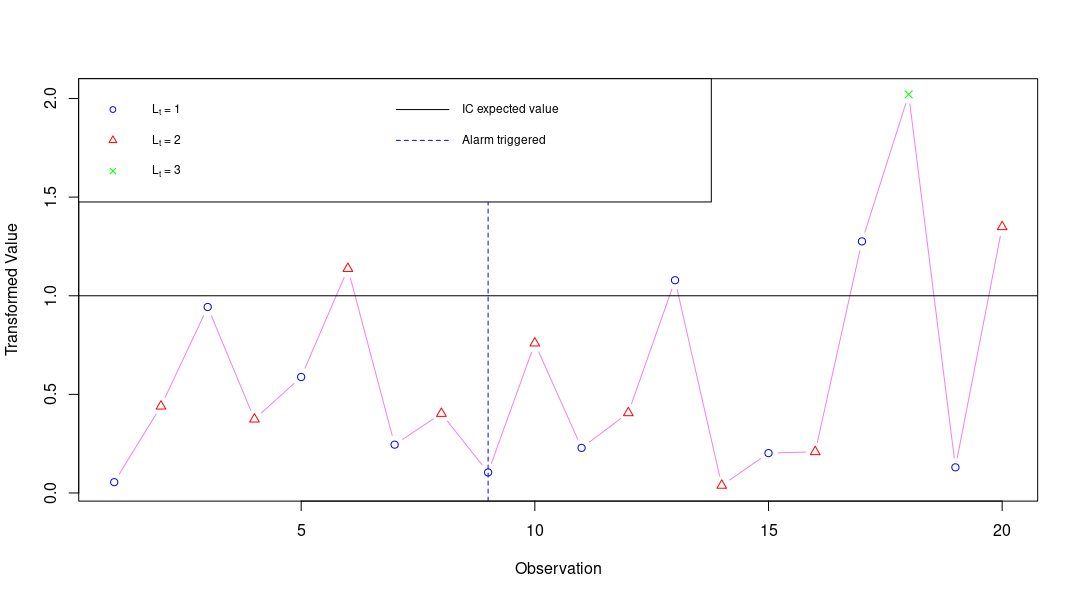}
\caption{The first 20 transformed observations $Z_t$, with color indicating the indicator value $L_t$ corresponding to each observation.}
\label{fig:transformeddata20}
\end{figure}

\begin{figure}[!htpb]
\centering
\includegraphics[width=.9\textwidth]{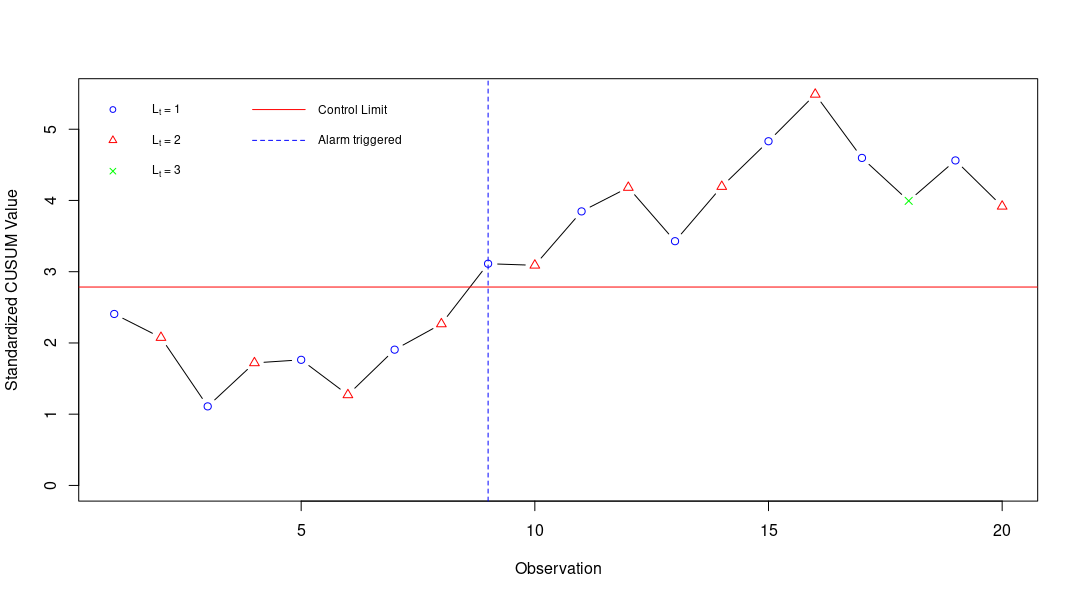}
\caption{Plot of $Q_t^{(+, +, +)}$ as a function of time along with the corresponding indicator value $L_t$. The solid red line denotes the control limit corresponding to a target $\text{ARL}_0$ of 50. For illustration, the CUSUM statistic continues to update beyond the alarm point to show how subsequent observations contribute to the chart.}
\label{fig:CUSUM}
\end{figure}

To better understand this behavior, we zoom in on the first 20 transformed observations $Z_t$, along with their corresponding indicator values $L_t$ (defined in~\eqref{eqn:Lt}), as shown in Figure~\ref{fig:transformeddata20}.
Among the first 9 observations, we see that---with the exception of the 6\textsuperscript{th}---all values lie below the IC mean, and none is associated with an indicator value of $L_t = 3$. This pattern suggests potential increases in both $k_1$ and $k_2$. Accordingly, we would expect either $Q_t^{(+, +, +)}$ or $Q_t^{(+, +, -)}$ to exceed the control limit at the time when the alarm is triggered. Since these two statistics track nearly identical values up to the alarm point, we plot only $Q_t^{(+, +, +)}$ for the first 20 observations in Figure~\ref{fig:CUSUM}. As shown in the figure, $Q_t^{(+, +, +)}$ indeed triggers the alarm at the 9\textsuperscript{th} observation, confirming potential increases in  both $k_1$ and $k_2$.

All the above plots underscore the practical advantages of our adaptive CUSUM chart over the Shewhart chart of \cite{Zwetsloot_MainPaper}. In particular, and consistent with our simulation findings, the adaptive CUSUM chart is substantially more effective at detecting decreases in the transformed mean (corresponding to increases in $k$), which in this context reflects shorter fault-free and repair times. While reductions in repair time may be of secondary importance in escalator maintenance, such changes are critical in many industrial applications where early detection of decreasing component lifetimes is essential.

Furthermore, even if a target $\text{ARL}_0 = 50$ is deemed overly aggressive, our adaptive CUSUM statistic surpasses a value of 5.5 by the 16\textsuperscript{th} observation---sufficient to trigger an alarm even under considerably more conservative settings (e.g., $\text{ARL}_0 \approx 3000$). In contrast, the Shewhart chart fails to detect any change even under the more liberal threshold of $\text{ARL}_0 = 50$, as it relies solely on isolated large deviations and lacks the ability to accumulate evidence over time. As a result, it struggles to detect subtle but persistent shifts in the absence of extreme values.

\section{Concluding Remarks}
In this paper, we propose an adaptive CUSUM procedure to monitor bivariate TBE data in real time. Our formulation follows that of \cite{Zwetsloot_MainPaper}, who developed a Shewhart chart for this monitoring task. While the Shewhart chart is known for its simplicity, it is generally inefficient at detecting small or moderate process shifts, as it relies solely on the most recent observation. Extending the Shewhart chart into a more efficient monitoring scheme---such as a CUSUM chart that incorporates historical information---introduces considerable complexity, as demonstrated in this paper. Despite these challenges, we have developed a principled approach that resolves each complication, ultimately yielding a fully adaptive CUSUM chart. Simulation results and our real-data application both demonstrate that the proposed adaptive CUSUM procedure offers substantial improvements in detection performance over the Shewhart chart of \cite{Zwetsloot_MainPaper}, particularly for persistent but moderate shifts in the underlying process parameters.

Despite these advantages, our method has some limitations. The primary appeal of the Shewhart chart lies in its computational simplicity. In contrast, our adaptive CUSUM chart requires estimating and storing the time-varying IC distributions of the charting statistics during the initial transient period before they converge to their stationary regime. This requirement can be both memory-intensive and computationally burdensome. Future work could explore analytic or semi-analytic approximations to these transient IC distributions, thereby reducing or eliminating the need to simulate and store empirical distributions. Another option is to initialize each adaptive CUSUM statistic and its associated estimators directly from their respective stationary (steady-state) distributions. This would ensure that all subsequent adaptive CUSUM statistics remain within the stationary regime, requiring storage only of their stationary IC distributions.

While our focus in this paper has been on bivariate TBE data, the proposed decorrelation and monitoring framework can, in principle, be extended to higher-dimensional multivariate TBE data. However, in higher dimensions, the OC distribution of the decorrelated observations becomes substantially more complex, as it depends on the temporal order of events across components. We plan to investigate this multivariate extension rigorously in future research.


\section*{Appendix: Proof} \label{sec:appendix}

To simplify notation, we omit the subscript \(i\) in all subsequent proofs, since the derivations apply to any observation \(\X_i\).

\noindent \textbf{Proof of Proposition \ref{prop:IC}:} 
We begin by deriving the IC distribution of \(U_{(1)}\).

\noindent \textbf{IC Distribution of $U_{(1)}$:}  Recall that
$U_{(1)} = F^{(0)}_{X_{(1)} \mid V}(X_{(1)} \mid V)$.
 For $u_{1} \in [0,1]$,
\[
\begin{aligned}
\mathbb{P}(U_{(1)} \leq u_1 \mid V) &= \mathbb{P}\left(F^{(0)}_{X_{(1)} \mid V}(X_{(1)} \mid V) \leq u_1 \mid V\right) \\
&= \mathbb{P}\left(X_{(1)} \leq {F^{(0)}}^{-1}_{X_{(1)} \mid V}(u_1 \mid V)\mid V\right) \\
&= F^{(0)}_{X_{(1)} \mid V}\left({F^{(0)}}^{-1}_{X_{(1)} \mid V}(u_1 \mid V) \mid V\right) = u_1,
\end{aligned}
\]
Therefore, $\mathbb{P}(U_{(1)} \leq u_1)=u_1$, indicating that 
 $U_{(1)} \sim \text{Unif}[0,1]$.

\noindent \textbf{IC Distribution of $U_{(2)}$:} Next, we derive the IC distribution of \(U_{(2)}\), where
$U_{(2)} = F^{(0)}_{X_{(2)} \mid X_{(1)}, V}(X_{(2)} \mid X_{(1)}, V)$.
For $u_2 \in [0,1]$,
\[
\begin{aligned}
\mathbb{P}(U_{(2)} \leq u_2 \mid X_{(1)}, V) &= \mathbb{P}\left(F^{(0)}_{X_{(2)}|X_{(1)},V}(X_{(2)} \mid X_{(1)}, V) \leq u_2 \mid X_{(1)}, V\right) \\
&= \mathbb{P}\left(X_{(2)} \leq {F^{(0)}}^{-1}_{X_{(2)}|X_{(1)},V}(u_2 \mid X_{(1)}, V)\mid X_{(1)}, V\right) \\
&=F^{(0)}_{X_{(2)}|X_{(1)},V}\left({F^{(0)}}^{-1}_{X_{(2)}|X_{(1)},V}(u_2 \mid X_{(1)}, V) \mid X_{(1)}, V\right) = u_2,
\end{aligned}
\]
Thus, $\mathbb{P}(U_{(2)} \leq u_2)=u_2$, indicating that 
 $U_{(2)} \sim \text{Unif}[0,1]$.

\noindent \textbf{Independence of \(\mathbf{U_{(1)}}\) and \(\mathbf{U_{(2)}}\):}  
From the derivation above, we observe that the conditional distribution of \(U_{(2)}\) given \(X_{(1)}\) and \(V \) does not depend on \(X_{(1)}\) and \(V\), implying that \(U_{(2)}\) and \((X_{(1)}, V)\) are independent. Since \(U_{(1)} = F^{(0)}_{X_{(1)} \mid V}(X_{(1)} \mid V)\) is a deterministic function of \((X_{(1)}, V)\), it follows that \(U_{(1)}\) and \(U_{(2)}\) are also independent.
\qed

\vspace{5mm}

\noindent \textbf{Proof of Proposition~\ref{prop:OC_MOBE}:}  
We first derive the OC distribution of \(U_{(1)}\).

\noindent \textbf{OC Distribution of \(\mathbf{U_{(1)}}\):}  
 Recall that
$U_{(1)} = F^{(0)}_{X_{(1)} \mid V}(X_{(1)} \mid V)=1 - \exp(-\Lambda X_{(1)})$.
Then, the CDF of \(U_{(1)}\) under the OC distribution is:
\[
\begin{aligned}
F_{U_{(1)}}(u) &= \mathbb{P}(U_{(1)} \leq u ) = \mathbb{P}\left(F^{(0)}_{X_{(1)} \mid V}(X_{(1)} \mid V) \leq u \right) \\
&= \mathbb{P}\left(1 - \exp(-\Lambda X_{(1)}) \leq u \right) = \mathbb{P}\left(X_{(1)} \leq -\frac{\log(1 - u)}{\Lambda} \right).
\end{aligned}
\]
Since \(\X \sim \text{MOBE}(\gamma_1, \gamma_2, \gamma_3)\), similar to how we derive (\ref{eqn:IC1_MOBE}), we can obtain the CDF of \(X_{(1)}\) as:
\[
F_{X_{(1)}}(x) = 1 - \exp(-\Gamma x), \quad \text{where } \Gamma := \gamma_1 + \gamma_2 + \gamma_3.
\]
Therefore,
\[
\begin{aligned}
F_{U_{(1)}}(u) &= \mathbb{P}\left(X_{(1)} \leq -\frac{\log(1 - u)}{\Lambda} \right) \\
&= 1 - \exp\left( -\Gamma \cdot \frac{-\log(1 - u)}{\Lambda} \right) \\
&= 1 - (1 - u)^{\Gamma / \Lambda}.
\end{aligned}
\]

\noindent \textbf{OC Distribution of \(\mathbf{U_{(2)}}\):}  
Next, we derive the OC distribution of \(U_{(2)}\), where $U_{(2)} = F^{(0)}_{X_{(2)} \mid X_{(1)}, V}(X_{(2)} \mid X_{(1)}, V)$.
Because $F^{(0)}_{X_{(2)}|X_{(1)},V}$ depends on \(V\), we consider each case separately.

\textbf{Case 1:} \(V = 0\)  
\[
\begin{aligned}
\mathbb{P}(U_{(2)} \leq u \mid X_{(1)}, V = 0) & = \mathbb{P}\left(F^{(0)}_{X_{(2)} \mid X_{(1)}, V}(X_{(2)} \mid X_{(1)}, 0) \leq u \mid X_{(1)} \right) \\
&= \mathbb{P}\left(1 - \exp[(\lambda_2 + \lambda_3)(X_{(1)} - X_{(2)})] \leq u \mid X_{(1)} \right) \\
&= \mathbb{P}\left(X_{(2)} \leq X_{(1)} - \frac{\log(1 - u)}{\lambda_2 + \lambda_3} \,\middle|\, X_{(1)} \right).
\end{aligned}
\]
Under the OC distribution \(\X \sim \text{MOBE}(\gamma_1, \gamma_2, \gamma_3)\), similar to how we derive (\ref{eqn:IC2_MOBE}), we can obtain the conditional distribution of \(X_{(2)} \mid X_{(1)}, V = 0\) as:
\[
F_{X_{(2)} \mid X_{(1)}}(y \mid x) = 1 - \exp[(\gamma_2 + \gamma_3)(x - y)].
\]
Therefore,
\[
\begin{aligned}
\mathbb{P}(U_{(2)} \leq u \mid X_{(1)}, V = 0) &= \mathbb{P}\left(X_{(2)} \leq X_{(1)} - \frac{\log(1 - u)}{\lambda_2 + \lambda_3} \,\middle|\, X_{(1)} \right) \\
&= 1 - \exp\left[(\gamma_2 + \gamma_3) \cdot \frac{\log(1 - u)}{\lambda_2 + \lambda_3} \right] \\
&= 1 - (1 - u)^{(\gamma_2 + \gamma_3)/(\lambda_2 + \lambda_3)},
\end{aligned}
\]
independent of $X_{(1)}$. As a result, $F_{U_{(2)}}(u) =  1 - (1 - u)^{(\gamma_2 + \gamma_3)/(\lambda_2 + \lambda_3)}$ when  \(V = 0\).  

\textbf{Case 2:} \(V = 1\)  
By symmetry, when \(V = 1\), the corresponding CDF is:
\[
F_{U_{(2)}}(u) = 1 - (1 - u)^{(\gamma_1 + \gamma_3)/(\lambda_1 + \lambda_3)}.
\]

\noindent \textbf{Independence of \(\mathbf{U_{(1)}}\) and \(\mathbf{U_{(2)}}\):}  
From the derivation above, we observe that the conditional distribution of \(U_{(2)}\) given \(X_{(1)}\) does not depend on \(X_{(1)}\), implying that \(U_{(2)}\) and \(X_{(1)}\) are independent. Since \(U_{(1)} = 1 - \exp(-\Lambda X_{(1)})\) is a deterministic function of \(X_{(1)}\), it follows that \(U_{(1)}\) and \(U_{(2)}\) are also independent.
\qed

\vspace{5mm}
 \noindent \textbf{Proof of Proposition \ref{prop:OC_MOBW}:} We begin by deriving the OC distribution of \(U_{(1)}\).

 \noindent \textbf{OC Distribution of $\mathbf{U_{(1)}}$:}
Recall that
\(
U_{(1)} = F^{(0)}_{X_{(1)} \mid V}(X_{(1)} \mid V)=1 - \exp(-\Lambda X_{(1)}^\eta).
\)
The CDF of $U_{(1)}$ under the OC distribution is: 
\[
\begin{aligned}
F_{U_{(1)}}(u) &= \mathbb{P}(U_{(1)} \leq u ) = \mathbb{P}\left(1 - \exp(-\Lambda X_{(1)}^\eta) \leq u \right) \\
&= \mathbb{P}\left(X_{(1)} \leq \left(-\frac{\log(1 - u)}{\Lambda} \right)^{1/\eta} \right).
\end{aligned}
\]
Since $\X \sim$ MOBW$(\gamma_1,\gamma_2,\gamma_3,\eta)$, similar to how we derive (\ref{eqn:IC1_MOBW}), we can obtain the CDF of $X_{(1)}$ as:
\[
F_{X_{(1)}}(x) = 1 - \exp(-\Gamma x^{\eta}). 
\]
Therefore,
\[
\begin{aligned}
F_{U_{(1)}}(u) &= \mathbb{P}\left(X_{(1)} \leq \left(-\frac{\log(1 - u)}{\Lambda} \right)^{1/\eta} \right)\\
&= 1 - \exp\left( -\Gamma \cdot \frac{-\log(1 - u)}{\Lambda} \right) \\
&= 1 - (1 - u)^{\frac{\Gamma}{\Lambda}}.
\end{aligned}
\]

\noindent \textbf{OC Distribution of $\mathbf{U_{(2)}}$:}
Here we have \(U_{(2)} = F^{(0)}_{X_{(2)}|X_{(1)},V}(X_{(2)} \mid X_{(1)}, V)\). Because $F^{(0)}_{X_{(2)}|X_{(1)},V}$ depends on $V$, we consider each case separately.

\textbf{Case 1:}  \(V = 0\) 
\[
\begin{aligned}
\mathbb{P}(U_{(2)} \leq u \mid X_{(1)}, V = 0) &= \mathbb{P}\left(F^{(0)}_{X_{(2)} \mid X_{(1)}, V}(X_{(2)} \mid X_{(1)}, V)\leq u \mid X_{(1)}, V=0\right) \\
&= \mathbb{P}\left( 1 - \exp[(\lambda_2 + \lambda_3) (X_{(1)}^{\eta}-X_{(2)}^{\eta})]  \leq u \mid X_{(1)} \right) \\
&= \mathbb{P}\left(X_{(2)} \leq \left( X_{(1)}^{\eta}-\frac{\log(1 - u)}{\lambda_2+\lambda_3}\right)^{1/\eta} \,\middle|\, X_{(1)} \right).
\end{aligned}
\]
Under the OC distribution $\X \sim$ MOBW$(\gamma_1,\gamma_2,\gamma_3,\eta)$, similar to how we derive (\ref{eqn:IC2_MOBW}), we can obtain the conditional CDF of $X_{(2)}$ given  $X_{(1)}$ and $V=0$ as:
\[
F_{X_{(2)}|X_{(1)}}(y|x) = 1 - \exp[(\gamma_2 + \gamma_3) (x^{\eta}-y^{\eta})].
\]
Therefore,
\[
\begin{aligned}
\mathbb{P}(U_{(2)} \leq u \mid X_{(1)}, V = 0)  &= \mathbb{P}\left(X_{(2)} \leq \left( X_{(1)}^{\eta}-\frac{\log(1 - u)}{\lambda_2+\lambda_3}\right)^{1/\eta} \,\middle|\, X_{(1)} \right)\\
&= 1 - \exp\left[(\gamma_2+\gamma_3) \cdot \frac{\log(1 - u)}{\lambda_2+\lambda_3} \right] \\
&= 1 - (1 - u)^{\frac{\gamma_2+\gamma_3}{\lambda_2+\lambda_3}},
\end{aligned}
\]
independent of independent of $X_{(1)}$. As a result, $F_{U_{(2)}}(u) =   1 - (1 - u)^{\frac{\gamma_2+\gamma_3}{\lambda_2+\lambda_3}}$ when  \(V = 0\).

\textbf{Case 2:} \(V = 1\) By symmetry, we find that in this case, the CDF of $U_{(2)}$ is given by : \[
F_{U_{(2)}}(u)  = 1 - (1 - u)^{\frac{\gamma_1 + \gamma_3}{\lambda_1 + \lambda_3}}
\]

 \noindent \textbf{Independence of \(\mathbf{U_{(1)}}\) and \(\mathbf{U_{(2)}}\):} From the derivation above, we observe that the conditional CDF of $U_{(2)}$ given $X_{(1)}$ does not depend on $X_{(1)}$ , implying that $U_{(2)}$ and $X_{(1)}$ are independent. Since $U_{(1)}=1 - \exp(-\Lambda X_{(1)}^{\eta})$, $U_{(2)}$ and $U_{(1)}$ are also independent.
 \qed

\vspace{5mm}
 \noindent \textbf{Proof of Proposition \ref{prop:OC_Gumbel}:} Recall that
\(
U_{(1)} = F^{(0)}_{X_{(1)} \mid V}(X_{(1)} \mid V)=1 - \exp(-[C(1,1)]^\delta X_{(1)}).
\)
The CDF of $U_{(1)}$ under the OC distribution is: 
\[
\begin{aligned}
F_{U_{(1)}}(u) &= \mathbb{P}(U_{(1)} \leq u ) = \mathbb{P}\left(1 - \exp(-[C(1,1)]^\delta X_{(1)}) \leq u \right) \\
&= \mathbb{P}\left(X_{(1)} \leq -\frac{\log(1 - u)}{[C(1,1)]^\delta}  \right).
\end{aligned}
\]
Since $\X \sim$ Gumbel$(\alpha_1,\alpha_2,\beta)$, similar to how we derive (\ref{eqn:IC1_Gumbel}), we can obtain the CDF of $X_{(1)}$ as:
\[
F_{X_{(1)}}(x) = 1 - \exp(-[D(1,1)]^\beta x).
\]
Therefore,
\[
\begin{aligned}
F_{U_{(1)}}(u) &= \mathbb{P}\left(X_{(1)} \leq -\frac{\log(1 - u)}{[C(1,1)]^\delta}  \right)\\
&= 1 - \exp\left( [D(1,1)]^\beta \cdot \frac{\log(1 - u)}{[C(1,1)]^\delta} \right) \\
&= 1 - (1 - u)^{[D(1,1)]^\beta/[C(1,1)]^\delta}.
\end{aligned}
\]
 \qed

 \vspace{5mm}
\noindent \textbf{Proof of Proposition \ref{prop:OC2_Gumbel}:} First, note that in the independent case where \(\delta = \beta = 1\), the IC conditional CDF, $F^{(0)}_{X_{(2)}|X_{(1)},V}$, in (\ref{eqn:IC2_Gumbel}) and (\ref{eqn:IC3_Gumbel}) simplifies to:
\[
F^{(0)}_{X_{(2)}|X_{(1)},V}(y \mid x, v) = 1 - \frac{\exp[-C(x,y)]}{ \exp[-C(1,1) \cdot x] } \quad \text{if } v = 0 
\]
\[
F^{(0)}_{X_{(2)}|X_{(1)},V}(y \mid x, v) = 1 - \frac{\exp[-C(y,x)]}{ \exp[-C(1,1) \cdot x] } \quad \text{if } v = 1
\]

\textbf{Case 1:} \(V = 0\)  
\[
\begin{aligned}
&\mathbb{P}(U_{(2)} \leq u \mid X_{(1)}, V = 0)\\
= &\mathbb{P}\left(1 - \frac{\exp[-C(X_{(1)},X_{(2)})]}{ \exp[-C(1,1) \cdot X_{(1)}] } \leq u \mid X_{(1)}, V = 0\right) \\
= & \mathbb{P}\left(X_{(2)} \leq -\theta_2 \cdot \ln(1 - u) + \theta_2 \cdot C(1,1) \cdot X_{(1)} - \frac{\theta_2}{\theta_1} X_{(1)} \mid X_{(1)}, V = 0\right)
\end{aligned}
\]
Since $\X \sim$ Gumbel$(\alpha_1,\alpha_2,\beta)$, similar to how we derive (\ref{eqn:IC2_Gumbel}), we can obtain the conditional CDF of $X_{(2)}$ given  $X_{(1)}$ and $V=0$ as:
\[
F_{X_{(2)}|X_{(1)},V}(y|x,V=0) = 1 - \frac{\exp[-D(x,y)]}{ \exp[-D(1,1) \cdot x] }.
\]
Therefore,
\[
\begin{aligned}
&\mathbb{P}(U_{(2)} \leq u \mid X_{(1)}, V = 0)\\
= & \mathbb{P}\left(X_{(2)} \leq -\theta_2 \cdot \ln(1 - u) + \theta_2 \cdot C(1,1) \cdot X_{(1)} - \frac{\theta_2}{\theta_1} X_{(1)}  \mid X_{(1)}, V = 0\right)\\
=&1 - \frac{\exp\left( -D(1,1) \cdot X_{(1)} + \frac{\theta_2}{\alpha_2} \cdot \ln(1 - u) \right)}{ \exp[-D(1,1) \cdot X_{(1)}] } \\
= & 1 - (1 - u)^{\theta_2 / \alpha_2},
\end{aligned}
\]
independent of $X_{(1)}$. As a result, $F_{U_{(2)}}(u) =  1 - (1 - u)^{\theta_2 / \alpha_2}$ when  \(V = 0\).  

\textbf{Case 2:} \(V = 1\) By symmetry, we can obtain the following CDF of $U_{(2)}$ when $V=1$:
\[
F_{U_{(2)} }(u) = 1 - (1 - u)^{\theta_1 / \alpha_1}.
\]

\noindent \textbf{Independence of \(\mathbf{U_{(1)}}\) and \(\mathbf{U_{(2)}}\):} From the derivation above, we observe that the conditional CDF of $U_{(2)}$ given $X_{(1)}$ does not depend on $X_{(1)}$, implying that $U_{(2)}$ and $X_{(1)}$ are independent. Since $U_{(1)}=1 - \exp(-[C(1,1)] X_{(1)})$, $U_{(2)}$ and $U_{(1)}$ are also independent.
 \qed

\section*{Acknowledgement}

The authors thank Inez Marie Zwetsloot for kindly sharing the escalator dataset used in the real-data application.

\section*{Disclosure of Interest}
No potential conflict of interest was reported by the authors.

\section*{Data Availability Statement}

The data used in this study originate from a collaboration between the City University of Hong Kong and a local company responsible for maintaining many of the city’s escalators. Further details are available in \cite{Zwetsloot_MainPaper} and \cite{HU2024}. 

\section*{Funding}

No funding was received to support this work.


\bibliographystyle{plainnat}
\bibliography{bibfile}

\begin{thebibliography}{15}
\providecommand{\natexlab}[1]{#1}
\providecommand{\url}[1]{\texttt{#1}}
\expandafter\ifx\csname urlstyle\endcsname\relax
  \providecommand{\doi}[1]{doi: #1}\else
  \providecommand{\doi}{doi: \begingroup \urlstyle{rm}\Url}\fi

\bibitem[Hu et~al.(2024)Hu, Qiu, Lin, Zwetsloot, Lee, Yeung, Yeung, and Wong]{HU2024}
Xuwen Hu, Jiaqi Qiu, Yu~Lin, Inez~Maria Zwetsloot, William Ka~Fai Lee, Edmond Yin~San Yeung, Colman Yiu~Wah Yeung, and Chris Chun~Long Wong.
\newblock An algorithm for modelling escalator fixed loss energy for phm and sustainable energy usage.
\newblock \emph{Energy and Buildings}, 314:\penalty0 114239, 2024.

\bibitem[Lee~Ho and Branco~Costa(2009)]{Costa_2008}
Linda Lee~Ho and Antonio~Fernando Branco~Costa.
\newblock Control charts for individual observations of a bivariate poisson process.
\newblock \emph{The International Journal of Advanced Manufacturing Technology}, 43:\penalty0 744--755, 2009.

\bibitem[Liu et~al.(2019)Liu, Zhang, and Mei]{Liu_etal2018}
Kun Liu, Ruizhi Zhang, and Yajun Mei.
\newblock Scalable sum-shrinkage schemes for distributed monitoring large-scale data streams.
\newblock \emph{Statistica Sinica}, 29\penalty0 (1):\penalty0 1--22, 2019.

\bibitem[Lorden and Pollak(2008)]{Lorden2008}
Gary Lorden and Moshe Pollak.
\newblock Sequential change-point detection procedures that are nearly optimal and computationally simple.
\newblock \emph{Sequential Analysis}, 27\penalty0 (4):\penalty0 476--512, November 2008.

\bibitem[Marshall and Olkin(1967)]{Marshall1967}
Albert~W. Marshall and Ingram Olkin.
\newblock A multivariate exponential distribution.
\newblock \emph{Journal of the American Statistical Association}, 62\penalty0 (317):\penalty0 30--44, 1967.

\bibitem[Moustakides(1986)]{Moustakides1986}
George~V. Moustakides.
\newblock Optimal stopping times for detecting changes in distributions.
\newblock \emph{The Annals of Statistics}, 14\penalty0 (4):\penalty0 1379--1387, 1986.

\bibitem[Ozsan et~al.(2010)Ozsan, Testik, and Weiß]{Ozsan2009}
Guney Ozsan, Murat~Caner Testik, and Christian~H. Weiß.
\newblock Properties of the exponential ewma chart with parameter estimation.
\newblock \emph{Quality and Reliability Engineering International}, 26\penalty0 (6):\penalty0 555–569, 2010.

\bibitem[Page(1954)]{Page1954}
E.~S. Page.
\newblock Continuous inspection schemes.
\newblock \emph{Biometrika}, 41\penalty0 (1):\penalty0 100--115, 1954.

\bibitem[Sparks et~al.(2019)Sparks, Jin, Karimi, Paris, and MacIntyre]{Sparks2019}
Ross Sparks, Brian Jin, Sarvnaz Karimi, Cecile Paris, and C.~R. MacIntyre.
\newblock Real-time monitoring of events applied to syndromic surveillance.
\newblock \emph{Quality Engineering}, 31\penalty0 (1):\penalty0 73–90, 2019.

\bibitem[Sparks et~al.(2020)Sparks, Joshi, Paris, Karimi, and MacIntyre]{Sparks2020}
Ross Sparks, Aditya Joshi, Cecile Paris, Sarvnaz Karimi, and C.~Raina MacIntyre.
\newblock Monitoring events with application to syndromic surveillance using social media data.
\newblock \emph{Engineering Reports}, 2\penalty0 (5), 2020.

\bibitem[Wu(2005)]{Wu_ChangepointEstim_2005}
Yanhong Wu.
\newblock \emph{Inference for Change-Point and Post-Change Means After a CUSUM Test}.
\newblock Springer New York, 2005.

\bibitem[Wu(2017)]{Wu2017}
Yanhong Wu.
\newblock Detecting changes in a multiparameter exponential family by using adaptive cusum procedure.
\newblock \emph{Sequential Analysis}, 36\penalty0 (4):\penalty0 467--480, 2017.

\bibitem[Xie et~al.(2002)Xie, Goh, and Kuralmani]{Xie.Textbook.2002}
M~Xie, T~N Goh, and V~Kuralmani.
\newblock \emph{Statistical Models and Control Charts for High-Quality Processes}.
\newblock Springer US, 2002.

\bibitem[Xie et~al.(2011)Xie, Xie, and Goh]{Xie_Goh_2011_TwoMEWMA_Gumbel}
Yujuan Xie, Min Xie, and Thong~Ngee Goh.
\newblock Two mewma charts for gumbel’s bivariate exponential distribution.
\newblock \emph{Journal of Quality Technology}, 43\penalty0 (1):\penalty0 50–65, 2011.

\bibitem[Zwetsloot et~al.(2023)Zwetsloot, Mahmood, Taiwo, and Wang]{Zwetsloot_MainPaper}
Inez~Maria Zwetsloot, Tahir Mahmood, Funmilola~Mary Taiwo, and Zezhong Wang.
\newblock A real‐time monitoring approach for bivariate event data.
\newblock \emph{Applied Stochastic Models in Business and Industry}, 39\penalty0 (6):\penalty0 789–817, 2023.

\end{thebibliography}

\end{document}